\documentclass[aps,pra,reprint,superscriptaddress,longbibliography,nofootinbib,floatfix]{revtex4-2}
\usepackage{amsmath,amssymb,bm,graphicx,booktabs}
\usepackage[T1]{fontenc}
\usepackage{placeins}
\usepackage{hyperref}
\hypersetup{colorlinks=true,citecolor=blue,linkcolor=blue,urlcolor=blue,pdftitle={Fast and High-Fidelity Motional-State-Preserving Transport in Optical Tweezers}}
\graphicspath{{figures/}}
\newcommand{\ket}[1]{\lvert #1\rangle}
\newcommand{\bra}[1]{\langle #1\rvert}
\newcommand{\avg}[1]{\langle #1\rangle}
\newcommand{\F}{\mathcal F}
\newcommand{\Uop}{\mathcal U}
\newcommand{\nbar}{\bar n}
\newcommand{\dd}{\mathrm d}
\makeatletter

\newenvironment{fixedcolumnfigure}{\begin{figure}[!htb]\centering\setlength{\abovecaptionskip}{5pt}}{\end{figure}}
\newenvironment{fixedcolumntable}{\par\addvspace{6pt}\noindent\begin{minipage}{\columnwidth}\def\@captype{table}\centering}{\end{minipage}\par\addvspace{6pt}}
\makeatother
\begin{document}
\title{Fast and High-Fidelity Motional-State-Preserving Transport in Optical Tweezers}
\author{Wen-hao Wang}
\affiliation{Department of Physics, Huazhong Normal University, Wuhan 430079, China}
\author{Jian-gong Hu}
\email{jghu@ncu.edu.cn}
\affiliation{School of Information Engineering, Nanchang University, Nanchang 330031, China}
\author{Guang-ming Huang}
\affiliation{Department of Physics, Huazhong Normal University, Wuhan 430079, China}
\author{Gao-xiang Li}
\email{gaox@ccnu.edu.cn}
\affiliation{Department of Physics, Huazhong Normal University, Wuhan 430079, China}
\author{Cheng Sheng}
\email{shengcheng@ccnu.edu.cn}
\affiliation{Department of Physics, Huazhong Normal University, Wuhan 430079, China}
\date{October 5, 2026}
\begin{abstract}
Reconfigurable neutral-atom processors require rapid, high-fidelity transport of atoms between optical-tweezer sites. In finite-depth Gaussian traps, fast motion can leave residual motional excitation even when the atom reaches its destination. We develop a basis-parameterized gradient-ascent pulse-engineering (GRAPE) framework for one-dimensional transport of a single $^{87}$Rb atom. A shortcut-to-adiabaticity reference is refined using 17 trajectory parameters, with ten additional parameters providing bounded trap-depth modulation; depth affects both confinement and the inverse construction of the physical trap center. For a $12.6\,\mu$m transfer completed in only $30.5\,\mu$s, corresponding to approximately $2.75$ periods of the reference trap, the selected joint protocol achieves a ground-state infidelity of $2.41\times10^{-7}$. Because preparing an exactly pure motional ground state is experimentally challenging, we also optimize population-weighted transport for initial distributions with mean vibrational occupations $\nbar\simeq0.16$ and $1.87$, reaching same-state probabilities of about $0.99993$ and $0.9934$. Trap-depth fluctuations can further degrade transport fidelity. Robust optimization under a measurement-informed depth-noise model improves the mean error, worst sampled error, and fidelity dispersion for the specified noise realizations. Coordinated control of tweezer position and confinement thus offers a route to rapid transport with low residual excitation for flexible atom rearrangement and dynamic connectivity in neutral-atom processors.
\end{abstract}
\maketitle

\FloatBarrier
\section{Introduction}
Optical-tweezer arrays have become programmable platforms for quantum computing and quantum simulation with neutral atoms \cite{Kaufman2021}. An optical tweezer confines an atom in the localized potential of a tightly focused laser beam; its vibrational states describe quantized motion within that potential \cite{Kaufman2021}. Individual atoms can be trapped, manipulated, and detected, while rearrangement of stochastic loading patterns enables the assembly of defect-free one- and two-dimensional target arrays \cite{Barredo2016,Endres2016}. Movable tweezers also allow qubit connectivity to change during a computation. Coherent transport of entangled atom arrays has demonstrated dynamic nonlocal connectivity, allowing distant qubits to be brought together for gates rather than restricting interactions to a fixed nearest-neighbor graph \cite{Bluvstein2022}. Reconfigurable logical processors combine this connectivity with spatially separated storage, entangling, and readout regions \cite{Bluvstein2024}. Beyond routing flexibility, atom rearrangement can implement the nonlocal syndrome-extraction circuits of high-rate quantum low-density parity-check codes; a proposed architecture exploits this capability to reduce the physical-qubit overhead of fault-tolerant computation \cite{Xu2024}. Thus effective any-to-any connectivity is an architectural resource, although its benefit depends on transport time, errors, and scheduling constraints. Rapid transport that preserves the motional conditions required by subsequent operations is consequently important for realizing these advantages.

Fast atomic transport has been studied theoretically through inverse engineering and shortcuts to adiabaticity (STA), which yield protocols with suppressed terminal vibrational excitation in harmonic models and can accommodate displacement and acceleration constraints or corrections beyond ideal harmonic dynamics \cite{Torrontegui2011,Chen2011,Ding2020,Guery2019}. Enhanced-STA studies have treated Gaussian optical traps and three-dimensional anharmonic lattice transport by evaluating dynamics beyond the idealized reference model \cite{Whitty2020,Hauck2021}. Related optical-transfer work has examined realistic trapping conditions \cite{Ness2018}. Complementary experiments have demonstrated nonadiabatic transport with low residual center-of-mass motion \cite{Couvert2008}, transport and depth control in optical microtrap arrays \cite{Schlosser2012}, and fast single-atom transport along straight and curved tweezer paths \cite{Hwang2025}. Optical conveyor-belt studies quantify how transport parameters affect heating and retention \cite{Hickman2020}, while site-resolved depth control in a fiber-array architecture enables repeated coherent qubit transfer \cite{Wang2026}. These approaches establish several routes to rapid transport, with distinct observables for motional excitation, retention, and coherence.

Preserving a motional quantum state imposes a stronger requirement than bringing the atomic center of mass to its destination. The terminal wave-packet width, phase profile, and vibrational populations must also be controlled. Slow transport suppresses nonadiabatic excitation at the cost of longer operation times and greater exposure to noise. For fast optical-tweezer transport, the finite depth and anharmonicity of the Gaussian potential become important as strong acceleration or higher initial vibrational occupation samples regions beyond the harmonic approximation. An analytical STA trajectory therefore provides a physically motivated reference that must be assessed in the full potential; Gaussian-trap corrections show why its fidelity need not carry over unchanged \cite{Whitty2020}. Recent work on transfer between tweezers has highlighted the role of static traps and coordinated control of the moving tweezer's position and depth \cite{Cicali2025}. A complementary question is how compact, explicitly coupled controls over trajectory and confinement can suppress residual excitation after fast translation in a fully time-dependent Gaussian potential.

Quantum optimal control provides a systematic way to refine a physical reference protocol \cite{Koch2022}. Chopped-basis methods such as CRAB and dCRAB provide compact control representations, and dCRAB has been applied to finite-temperature optical-tweezer transport \cite{Caneva2011,Pagano2024}. However, fixed-basis CRAB can encounter artificial local extrema introduced by basis truncation, whereas dCRAB mitigates this limitation by refreshing random search directions in successive superiterations \cite{Rach2015}. For model-based transport, gradient ascent pulse engineering (GRAPE) computes control derivatives efficiently through forward and adjoint propagation \cite{Koch2022}. In a conventional formulation where the control amplitudes in individual time slices are optimized independently, refining the control grid increases the number of search variables. Gradient optimization using parametrization (GROUP) demonstrates how gradient-based updates can instead be combined with a reduced control basis \cite{Sorensen2018}. We adopt this principle in a basis-parameterized GRAPE search using smooth control functions, allowing the propagation resolution to be refined without increasing the number of optimization variables. The basis can enforce endpoint conditions, although a finite expansion alone does not impose a hard actuator bandwidth limit.

Raman sideband cooling has prepared individual atoms in optical tweezers with high three-dimensional motional ground-state occupation \cite{Kaufman2012,WangCooling2019,Spence2022,Holzl2023}. In particular, single-$^{87}$Rb experiments demonstrate experimentally accessible near-ground-state preparation \cite{Kaufman2012,Spence2022}. Cooling nevertheless leaves finite excited-state populations, and a three-dimensional ground-state fraction does not uniquely determine a one-dimensional thermal distribution. We therefore use $\nbar\simeq0.16$ as a representative near-ground-state initial condition rather than attributing that exact value to a cited experiment, and also consider the broader $\nbar\simeq1.87$ case. Occupied states differ in spatial extent, anharmonic energy shifts, and nonadiabatic response, so a ground-state-optimized waveform need not preserve a mixture. Prior finite-temperature tweezer-transport work has also examined optimal control and laser-noise effects \cite{Pagano2024}. Our objective weights each occupied state's probability of returning to its corresponding final vibrational state; it differs from atom retention or preservation of arbitrary coherent superpositions.

Beyond the initial-state distribution, technical fluctuations can further limit high-fidelity transport. Spring-constant and optical-lattice noise studies show that transport sensitivity depends on the fluctuation spectrum and the chosen trajectory \cite{Lu2018,Lu2020}; general quantum-control analyses likewise connect robustness to noise statistics and pulse design \cite{Hocker2014}. We focus on depth noise motivated by laboratory observations of slow optical-power drift and narrowband spectral components. These observations inform a specified noise model, whose amplitudes are modeling assumptions rather than a fully calibrated optical-noise spectrum. Robust-control studies use several distinct formulations for uncertain dynamics, including trajectory optimization and disorder-averaged evolution \cite{Propson2022,Araki2023}. Robust optimization must balance nominal performance with average performance and the unfavorable tail of that ensemble. Treating initial-state populations and depth fluctuations explicitly allows their effects to be evaluated separately from propagation discretization and control-space truncation.

In this work, we construct a basis-parameterized GRAPE framework for fast one-dimensional transport of a single $^{87}$Rb atom in a Gaussian optical tweezer. All state evolution uses the full time-dependent Gaussian Hamiltonian in the laboratory frame. The local harmonic approximation supplies a minimum-jerk auxiliary reference and the depth--frequency relation used in the inverse construction of the physical trap center. Ten auxiliary-trajectory coefficients and seven residual center corrections give 17 trajectory variables; adding ten bounded depth coefficients yields 27 joint-control variables. At fixed waist, $\omega^2(t)=4U(t)/(mw_0^2)$, so depth modulation changes both the potential and the commanded center through the inverse construction. The corresponding coefficient gradients retain both contributions. The auxiliary trajectory is not assumed to equal the centroid in the full Gaussian dynamics.

The ground-state study compares fixed-depth trajectory optimization with joint center--depth optimization through control waveforms, resolved final-state populations, and stage-resolved search histories. At the longer demonstrated duration, trajectory optimization already reaches the $10^{-7}$ infidelity scale; the separate $28.5\,\mu$s task tests the additional depth freedom at a shorter duration. We then optimize a common waveform for initial distributions with mean vibrational occupations $\nbar\simeq0.16$ and $1.87$, obtaining population-weighted same-state transport probabilities of approximately $0.99993$ and $0.9934$, respectively. A complementary mixed-state fidelity checks agreement of the final density operator with its thermal target. Finally, the specified depth-noise model is incorporated into a robust objective and evaluated on a paired 192-scenario ensemble. Terminal fidelities used for quantitative comparisons are independently verified using refined numerical propagation; numerical settings, force diagnostics, and their limitations are collected in the appendices.

\begin{figure*}[!t]
\centering
\includegraphics[width=0.94\textwidth]{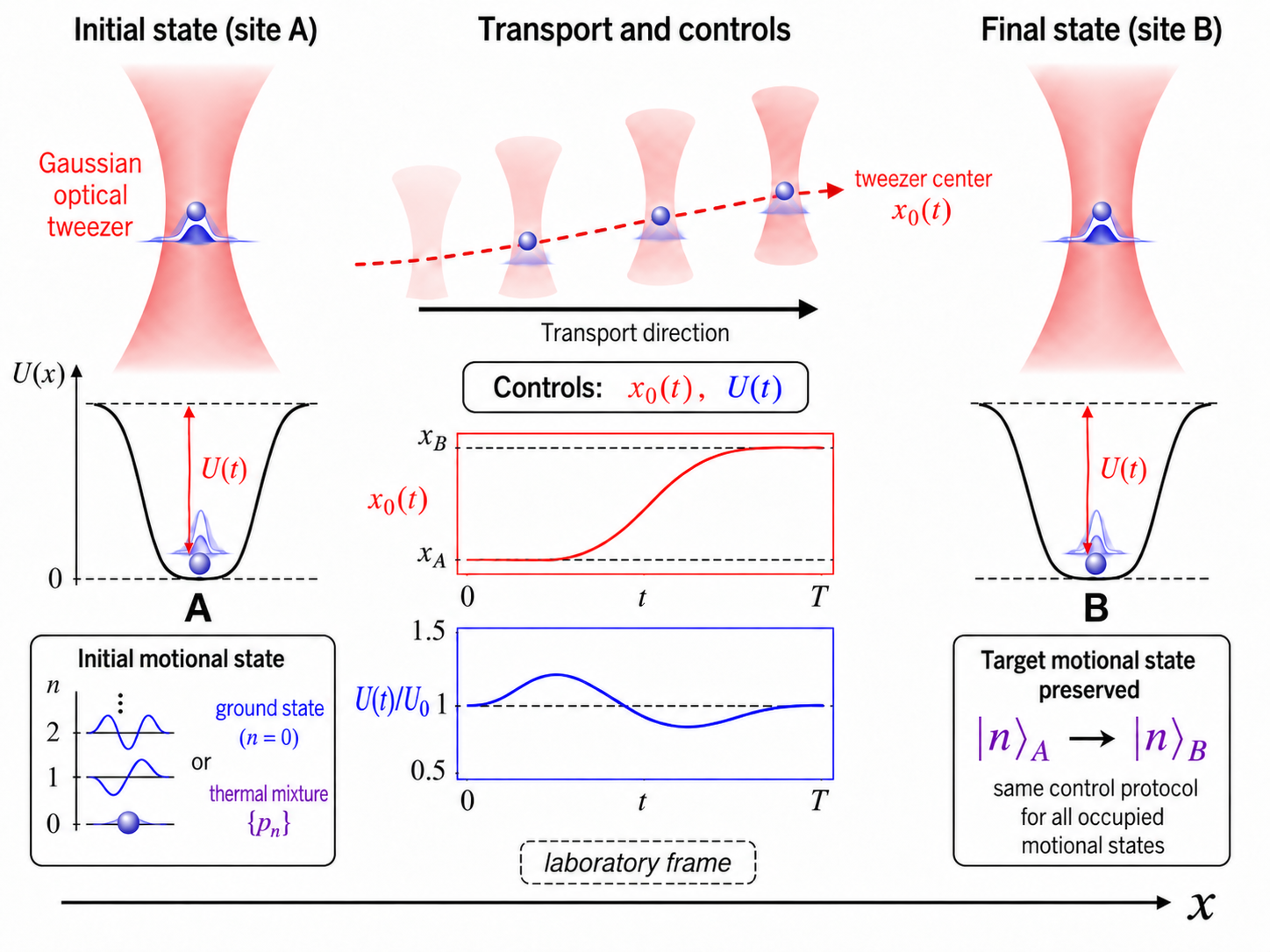}
\caption{Single-atom transport and control objective. A Gaussian optical tweezer carries the atom from site A to site B with center $x_0(t)$ and depth $U(t)$. The initial state is either the motional ground state or an incoherent mixture with populations $p_n$. A common waveform is used to preserve the occupied motional-state labels. Here $T$ denotes $t_f$. The potential-energy zero and illustrative depth excursion are schematic, whereas Eq.~\eqref{eq:H} uses zero potential at infinity and the numerical depth modulation is bounded by $5\%$.}
\label{fig:schematic}
\end{figure*}

Section II presents the Gaussian model, STA reference, parameterized controls, and staged optimization. Section III analyzes ground-state transport, joint control, and changes in transport conditions. Section IV extends the control to finite motional populations, and Sec. V examines robustness to depth noise. Section VI discusses the conclusions and limitations. The appendices give numerical checks, derivative verification, dynamical diagnostics, force bounds, and the noise-sampling protocol.

\FloatBarrier
\section{Model and parameterized optimal control}
\subsection{Gaussian tweezer and reference transport}
We consider a single atom of mass $m$ moving along the transport coordinate $x$. Its one-dimensional Hamiltonian is \cite{Pagano2024}
\begin{equation}
 H(t)=\frac{p^2}{2m}-U(t)\exp\!\left[-\frac{2[x-x_0(t)]^2}{w_0^2}\right],
 \label{eq:H}
\end{equation}
where $p=-i\hbar\partial_x$ is the momentum operator, $\hbar$ is the reduced Planck constant, $U(t)>0$ is the trap depth, $w_0$ is the fixed waist, and $x_0(t)$ is the physical trap center. The trap moves from $x_A=0$ to $x_B=L$ in time $t_f$. Figure~\ref{fig:schematic} summarizes the controls and state-preservation objective. The baseline parameters are listed in Table~\ref{tab:physical}. At fixed $w_0$, changing $x_0$ translates the potential, while changing $U$ changes its strength. In Table~\ref{tab:physical}, $k_B$ is Boltzmann's constant and $U_0/k_B$ expresses the trap-depth energy in temperature units; the initial motional temperature is specified separately. Transverse dynamics, photon scattering, and internal-state decoherence are outside this one-dimensional model.

Expansion about the trap minimum gives
\begin{equation}
 \omega^2(t)=\frac{4U(t)}{mw_0^2}.
 \label{eq:omega}
\end{equation}
This relation is used to construct controls, not to replace the propagation Hamiltonian. With $s=t/t_f$, the minimum-jerk reference is \cite{Cicali2025}
\begin{equation}
 q_{\rm ref}(t)=L(10s^3-15s^4+6s^5).
 \label{eq:qref}
\end{equation}
It satisfies prescribed positions and zero velocity and acceleration at both endpoints. For constant depth, the harmonic reference coordinate obeys $\ddot q_{\rm ref}+\omega_0^2(q_{\rm ref}-x_0)=0$. Solving for the center gives the harmonic inverse construction \cite{Torrontegui2011,Chen2011}
\begin{equation}
 x_0^{\rm STA}(t)=q_{\rm ref}(t)+\frac{\ddot q_{\rm ref}(t)}{\omega_0^2}.
 \label{eq:STA}
\end{equation}
Here $q_{\rm ref}$ is an auxiliary coordinate. It is not imposed as the actual centroid in Gaussian propagation.

\begin{fixedcolumntable}
\caption{Baseline physical parameters. Distance and duration scans change only the indicated transport task. The waist is fixed by the depth and reference frequency; the displayed value is rounded.}
\label{tab:physical}
\begin{ruledtabular}
\begin{tabular}{lll}
Quantity & Symbol & Value\\
\hline
Atomic mass & $m$ & $1.44316060\times10^{-25}$ kg\\
Distance & $L$ & $12.6\,\mu$m\\
Duration & $t_f$ & $30.5\,\mu$s\\
Reference depth & $U_0/k_B$ & $0.8$ mK\\
Reference frequency & $\omega_0/(2\pi)$ & $90$ kHz\\
Waist & $w_0$ & $0.978446\,\mu$m
\end{tabular}
\end{ruledtabular}
\end{fixedcolumntable}

\subsection{Smooth control basis and boundary conditions}
The auxiliary trajectory and residual center correction are
\begin{align}
 q_c(t)&=q_{\rm ref}(t)+L\sum_{k=1}^{10}a_k B_k(s),\label{eq:qc}\\
 \delta x_{\rm res}(t)&=L\sum_{j=1}^{7}r_jR_j(s),\qquad R_j=B_{j+3},\label{eq:res}\\
 B_k(s)&=\frac{s^3(s-1)^3\sin(k\pi s)}{N_k^B},\label{eq:B}
\end{align}
with $N_k^B=\max_{0\le s\le1}|s^3(s-1)^3\sin(k\pi s)|$.
Each $B_k$ and its first two derivatives vanish at the endpoints. The physical center for trajectory-only optimization is
\begin{equation}
 x_0(t)=q_c(t)+\frac{\ddot q_c(t)}{\omega_0^2}+\delta x_{\rm res}(t).
 \label{eq:traj}
\end{equation}
The 10 auxiliary and 7 residual coefficients supply 17 control variables. The auxiliary block deforms the transport coordinate before inverse construction, whereas the residual block acts directly on the physical center. The second derivative in Eq.~\eqref{eq:traj} gives higher auxiliary modes increasingly strong acceleration-weighted contributions. Restricting the residual block to modes 4--10 provides complementary direct-center flexibility. The two blocks need not be orthogonal.

To release the confinement strength, we use
\begin{align}
 U(t)&=U_0[1+\epsilon_U\tanh z(s)],\quad \epsilon_U=0.05,\label{eq:depth}\\
 z(s)&=\sum_{k=1}^{10}c_kD_k(s),\qquad
 D_k(s)=\frac{\sin^2(\pi s)\sin(k\pi s)}{N_k^D},\label{eq:D}
\end{align}
where $N_k^D$ normalizes the maximum absolute amplitude. This ensures $0.95U_0<U(t)<1.05U_0$ and $U(0)=U(t_f)=U_0$, with vanishing first and second endpoint derivatives. The joint physical center is
\begin{equation}
 x_0(t)=q_c(t)+\frac{\ddot q_c(t)}{\omega^2(t)}+\delta x_{\rm res}(t).
 \label{eq:center}
\end{equation}
The joint vector $\bm\theta=(\bm a,\bm c,\bm r)$ has 27 entries. The bounds are $|a_k|,|r_j|\le2\times10^{-3}$ and $|c_k|\le1.5$.

The first two terms in Eq.~\eqref{eq:center} retain the harmonic translation construction with an instantaneous frequency, while $\delta x_{\rm res}$ adds a direct center correction. This relation alone does not impose independent conditions on wave-packet width when confinement changes; final-state recovery is determined by optimizing the Gaussian dynamics \cite{Torrontegui2011}.

These conditions preserve the endpoint \emph{positions} of the physical center and the reference endpoint spectrum. They do not impose zero endpoint velocity and acceleration on $x_0$ itself: derivatives of $\ddot q_c/\omega^2$ also enter those quantities. Likewise, a bound on an individual auxiliary coefficient does not directly bound its total physical-center contribution, because the differentiated basis appears in Eq.~\eqref{eq:center}. The parameterization supplies smooth finite-dimensional controls, but is not a hard experimental bandwidth or slew-rate constraint.

\subsection{Objectives, coupled gradients, and staged optimization}
Let $\ket{\phi_n^A}$ and $\ket{\phi_n^B}$ denote corresponding eigenstates of the initial and final static Gaussian traps. The index $n$ orders the bound states by energy, starting with the ground state $n=0$. Both endpoint traps have depth $U_0$ and waist $w_0$, so corresponding states have the same spectrum and differ by spatial translation. The operator $\Uop(t_f,0)$ denotes evolution under Eq.~\eqref{eq:H}. Ground-state fidelity is \cite{Cicali2025}
\begin{equation}
 \F_0=\left|\bra{\phi_0^B}\Uop(t_f,0)\ket{\phi_0^A}\right|^2,
 \qquad J=-\ln\F_0.
 \label{eq:F0}
\end{equation}
Near unit fidelity, $J\simeq1-\F_0$. The logarithm gives a monotone loss with a convenient high-fidelity scale. We report infidelity rather than an optimizer-rescaled loss. The objective tests the complete final state at $t_f$, up to a global phase, and allows transient excitation during transport.

The propagation resolution and control-space dimension are independent: thousands of time steps resolve the wave function, while only 17 or 27 smooth coefficients are optimized. Let $t_\ell=\ell\Delta t$, $\Delta t=t_f/M$, and let $V_\ell$ denote the Gaussian potential of Eq.~\eqref{eq:H} at the midpoint $t_{\ell+1/2}$. With $T=p^2/(2m)$, forward propagation uses the second-order split step
\begin{equation}
\begin{aligned}
 \ket{\psi_{\ell+1}}&=S_\ell\ket{\psi_\ell},
 &\ket{\psi_0}&=\ket{\phi_0^A},\\
 S_\ell&=P_\ell K P_\ell,
 &P_\ell&=e^{-iV_\ell\Delta t/(2\hbar)},\quad
 K=e^{-iT\Delta t/\hbar}.
\end{aligned}
\label{eq:forward}
\end{equation}
The kinetic factor $K$ is applied in Fourier space. The forward pass gives $\ket{\psi_M}$, its overlap $A=\langle\phi_0^B|\psi_M\rangle$, and $J=-\ln|A|^2$.

The adjoint state carries the sensitivity of the final error backward through the dynamics. Pairing it with the forward state determines how a control change at each time affects that error.

For $\F_0=|A|^2>0$, choose the terminal adjoint $\ket{\lambda_M}=-(A/\F_0)\ket{\phi_0^B}$, so that $\delta J=2\operatorname{Re}\langle\lambda_M|\delta\psi_M\rangle$. It is propagated backward by $\ket{\lambda_\ell}=S_\ell^\dagger\ket{\lambda_{\ell+1}}$. The forward and adjoint states give the time-slice derivatives
\begin{equation}
 g_{q,\ell}=2\operatorname{Re}\bra{\lambda_{\ell+1}}
 \frac{\partial S_\ell}{\partial q_\ell}\ket{\psi_\ell},
 \qquad q_\ell\in\{x_{0,\ell},U_\ell\}.
 \label{eq:timegrad}
\end{equation}
Thus $g_{x,\ell}=\partial J/\partial x_{0,\ell}$ and $g_{U,\ell}=\partial J/\partial U_\ell$ are derivatives of the discretely propagated Gaussian model with respect to independent midpoint controls. The derivative of $S_\ell$ acts on both potential half steps; the kinetic factor has no direct control dependence. This forward--adjoint calculation is the GRAPE gradient step \cite{Koch2022,deFouquieres2011}. Rather than updating every midpoint control independently, we project these gradients onto the basis coefficients, as in gradient-based control with parameterized waveforms \cite{Sorensen2018}:
\begin{equation}
 \frac{\partial J}{\partial\theta_\alpha}
 =\sum_\ell\left[
 g_{x,\ell}\frac{\partial x_{0,\ell}}{\partial\theta_\alpha}
 +g_{U,\ell}\frac{\partial U_\ell}{\partial\theta_\alpha}
 \right].
 \label{eq:chain}
\end{equation}
The local derivatives already contain the propagation time step through $\partial S_\ell/\partial q_\ell$, so no additional $\Delta t$ multiplies this sum. With primes denoting derivatives with respect to $s$,
\begin{align}
 \frac{\partial x_0}{\partial a_k}
 &=L\left[B_k+\frac{B_k''}{t_f^2\omega^2(t)}\right],
 &\frac{\partial x_0}{\partial r_j}&=LR_j,\label{eq:ja}\\
 \frac{\partial U}{\partial c_k}
 &=\epsilon_U U_0\operatorname{sech}^2z\,D_k,\label{eq:jc1}\\
 \frac{\partial x_0}{\partial c_k}
 &=-\frac{mw_0^2\ddot q_c}{4U^2}\frac{\partial U}{\partial c_k}.
 \label{eq:jc2}
\end{align}
Omitting Eq.~\eqref{eq:jc2} would differentiate a different control construction.

L-BFGS-B does not calculate the gradient: it uses the GRAPE-derived coefficient gradient to update the finite-dimensional parameter vector under the bounds specified above \cite{MoralesNocedal2011}. The search uses scaled coordinates $z_\alpha=\theta_\alpha/s_\alpha$, where $s_\alpha$ is the bound of the corresponding coefficient block. This removes trivial scale differences between trajectory and depth parameters without changing the physical controls. The scaled gradient obeys $\partial J/\partial z_\alpha=s_\alpha\,\partial J/\partial\theta_\alpha$. The vector $\bm z$ is distinct from the scalar depth function $z(s)$. At each iteration L-BFGS-B uses recent gradient differences to approximate local curvature, constructs a bound-feasible search direction, and selects a step by line search. Each trial coefficient vector reconstructs $q_c(t)$, $U(t)$, and the coupled physical center $x_0(t)$; these determine the Hamiltonian, the forward and adjoint propagation, and the next objective and gradient evaluation. Appendix~\ref{app:basis} gives the optimizer settings.

The joint search starts with all coefficients zero, which exactly reproduces the STA reference. Stage 1 optimizes $(\bm a,\bm r)$ at fixed depth $U(t)=U_0$; stage 2 holds these coefficients fixed and releases $\bm c$; stage 3 releases all coefficients for joint refinement. During stage 2 the physical center still changes through Eq.~\eqref{eq:center}. The procedure is therefore depth-\emph{coupled} refinement, rather than fixed-center depth optimization. These stages organize a nonconvex search and are not a guarantee of global optimality or monotonically improving independently validated fidelity.

\subsection{Propagation and validation}
All state evolution uses Eq.~\eqref{eq:H} in the laboratory frame, with Fourier kinetic propagation and midpoint second-order split-operator evolution \cite{Gharibnejad2020}. Independent validation reconstructs each control from its fixed coefficients and repropagates the state without reoptimization. Appendix~\ref{app:numerics} specifies the spatial domain, discretization tests, and initial-state truncation for the reported comparisons. Appendix~\ref{app:gradient} verifies the discrete gradient independently. Gradient verification checks the derivative of the discrete model; propagation refinement checks sensitivity to numerical resolution; control-space comparisons assess how the optimized result depends on the available waveform freedom.

\FloatBarrier
\section{Ground-state transport}
\setlength{\textfloatsep}{10pt plus 2pt minus 2pt}
\setlength{\intextsep}{8pt plus 2pt minus 2pt}
\setlength{\floatsep}{8pt plus 2pt minus 2pt}
\setlength{\dbltextfloatsep}{12pt plus 2pt minus 2pt}
\begin{figure*}[!t]
\centering
\includegraphics[width=0.97\textwidth]{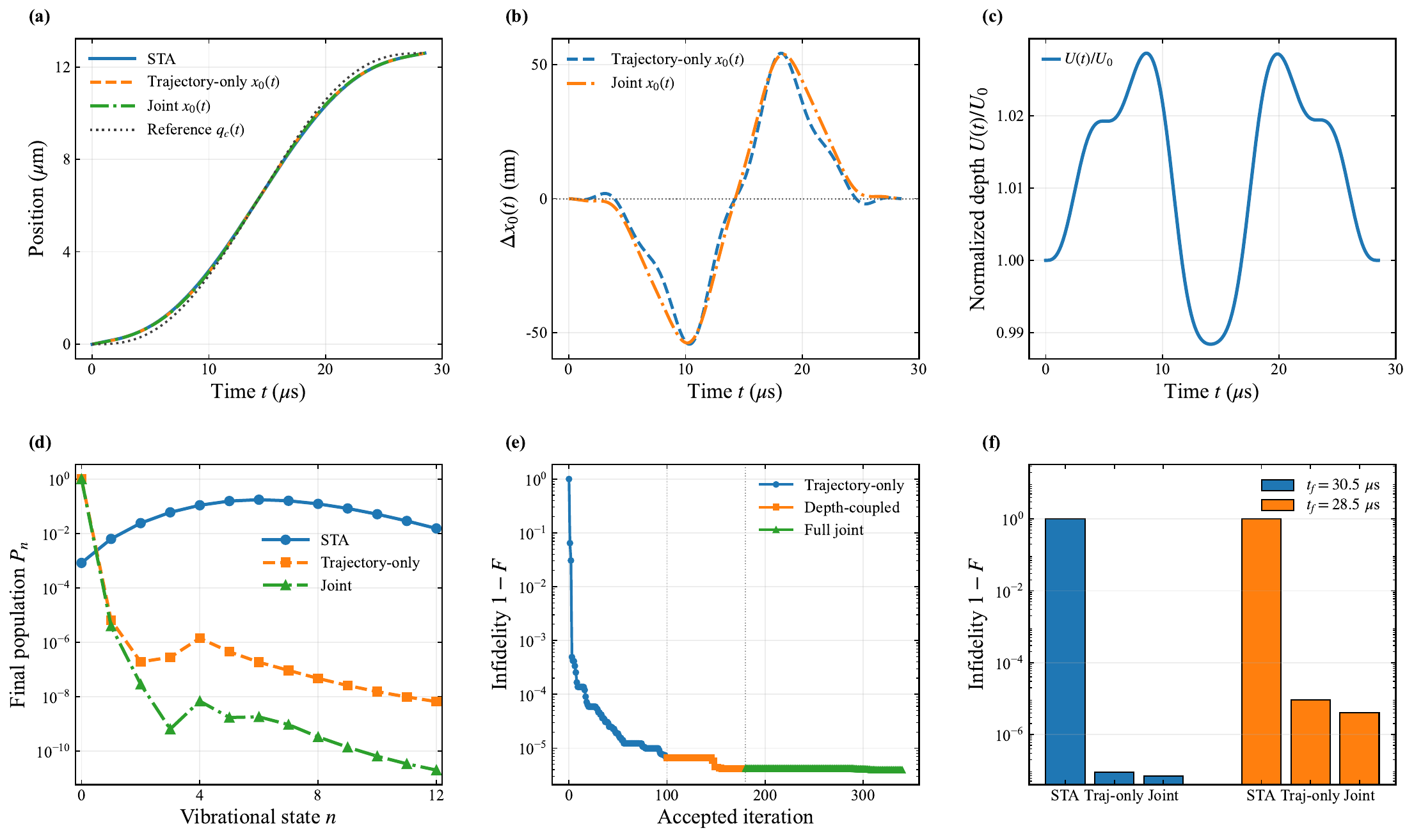}
\caption{Ground-state transport over $L=12.6\,\mu$m. (a)--(e) use $t_f=28.5\,\mu$s: (a) physical trap centers for STA, trajectory-only, and joint protocols, with the auxiliary $q_c(t)$; (b) optimized center corrections relative to STA; (c) joint depth $U(t)/U_0$; (d) terminal populations $P_n$ for the displayed vibrational states; (e) infidelity across accepted iterations of the trajectory-only, depth-coupled, and full-joint stages; vertical guides mark the stage boundaries. (f) Terminal infidelities for separately optimized $30.5$- and $28.5\,\mu$s tasks after refined repropagation. The STA bars are shown at unity scale.}
\label{fig:joint}
\end{figure*}
\subsection{Control structure and suppression of final excitation}
The harmonic STA construction supplies an organized starting point, but its endpoint conditions do not ensure recovery of the target ground state in the full Gaussian potential. The need to evaluate a control in the physical potential, rather than only in its reference approximation, is shared with enhanced-STA work on Gaussian and anharmonic traps \cite{Whitty2020,Hauck2021}. Trajectory optimization adjusts the physical drive while leaving the translated spectrum fixed. Joint optimization additionally changes confinement and its time-dependent spectrum, thereby expanding the accessible wave-function dynamics.

Figure~\ref{fig:joint} organizes this comparison at a common distance $L=12.6\,\mu$m. Panels (a)--(e) concern the faster $t_f=28.5\,\mu$s task: controls, resolved final populations, and stage-resolved search histories. Panel (f) compares refined terminal errors at $t_f=30.5\,\mu$s and $28.5\,\mu$s. In panel (b), the center correction is
\begin{equation}
 \Delta x_0(t)=x_{0,\rm opt}(t)-x_{0,\rm STA}(t),
\end{equation}
with ``opt'' denoting either optimized protocol. For the terminal wave function, define the target-trap populations by
\begin{equation}
 P_n=|\bra{\phi_n^B}\psi(t_f)\rangle|^2.
 \label{eq:cumulativeexc}
\end{equation}
Here $n$ is the vibrational-state label and $P_0=\F_0$. The total ground-state infidelity is $\varepsilon=1-P_0$. The displayed excited-state populations do not exhaust $\varepsilon$, because higher bound states and continuum components may also contribute.

The physical center waveforms nearly coincide on the micrometer scale, while panel (b) resolves corrections of tens of nanometers. The $28.5\,\mu$s joint depth in panel (c) remains within the permitted $5\%$ range. Thus the gain comes from adjusting the drive and confinement during evolution while preserving the overall transport path. The auxiliary $q_c$ is a design coordinate rather than an assumed exact centroid in the Gaussian potential; Appendix~\ref{app:force} examines this distinction directly.

Panel (d) resolves individual final populations on a logarithmic scale. The optimized controls suppress the large excited-state population left by STA, and joint control generally lowers the displayed populations relative to trajectory-only control. The nonmonotonic variation across $n$ reflects redistribution among channels; these finite-state projections alone do not reconstruct the entire terminal infidelity.

Panel (e) shows the accepted-iteration histories of the faster-task staged search. The stage boundaries occur at cumulative iterations 100 and 180, and the full-joint stage ends at iteration 340. Trajectory optimization removes most of the STA error; subsequent depth-coupled and full-joint stages further lower the selected error. Changing depth in the second stage also changes the physical center through Eq.~\eqref{eq:center}. The history does not imply a global optimum.

In the refined terminal comparison of panel (f), the longer-duration trajectory and joint errors are $2.93\times10^{-7}$ and $2.41\times10^{-7}$; the selected trajectory-only candidate already performs well. The $28.5\,\mu$s comparison also favors the selected joint control, with both optimized protocols far below the STA error. This demonstrates the usefulness of the enlarged control space for these tasks, without claiming superiority over every trajectory-only iterate or initialization. Control-basis sensitivity and candidate-selection details are given in Appendix~\ref{app:basis}.

A high terminal fidelity does not require the atom to remain in the instantaneous ground state throughout transport: excitation amplitudes generated at different times can interfere on arrival. For the $30.5\,\mu$s task, Appendix~\ref{app:mechanism} examines changes in transition couplings and accumulated spectral phases, and tests the terminal sensitivity to local removal of the optimized depth increment.

\subsection{Changes in transport distance and duration}
To test the same construction under more demanding transport conditions, Fig.~\ref{fig:scan} varies distance at fixed duration and duration at fixed distance. For a fixed normalized reference shape, the acceleration scale grows as $L/t_f^2$. The Gaussian restoring force reaches a maximum and then weakens away from the center, so these scans probe increasingly demanding acceleration and final-state recovery. The finite Gaussian restoring force provides a natural scale for interpreting these changes; displacement-bounded harmonic protocols provide a related, but model-dependent, transport criterion \cite{Chen2011,Ding2020}. Its relation to auxiliary acceleration and the actual wave-packet force is developed in Appendix~\ref{app:force}.
\begin{fixedcolumnfigure}
\includegraphics[width=\columnwidth]{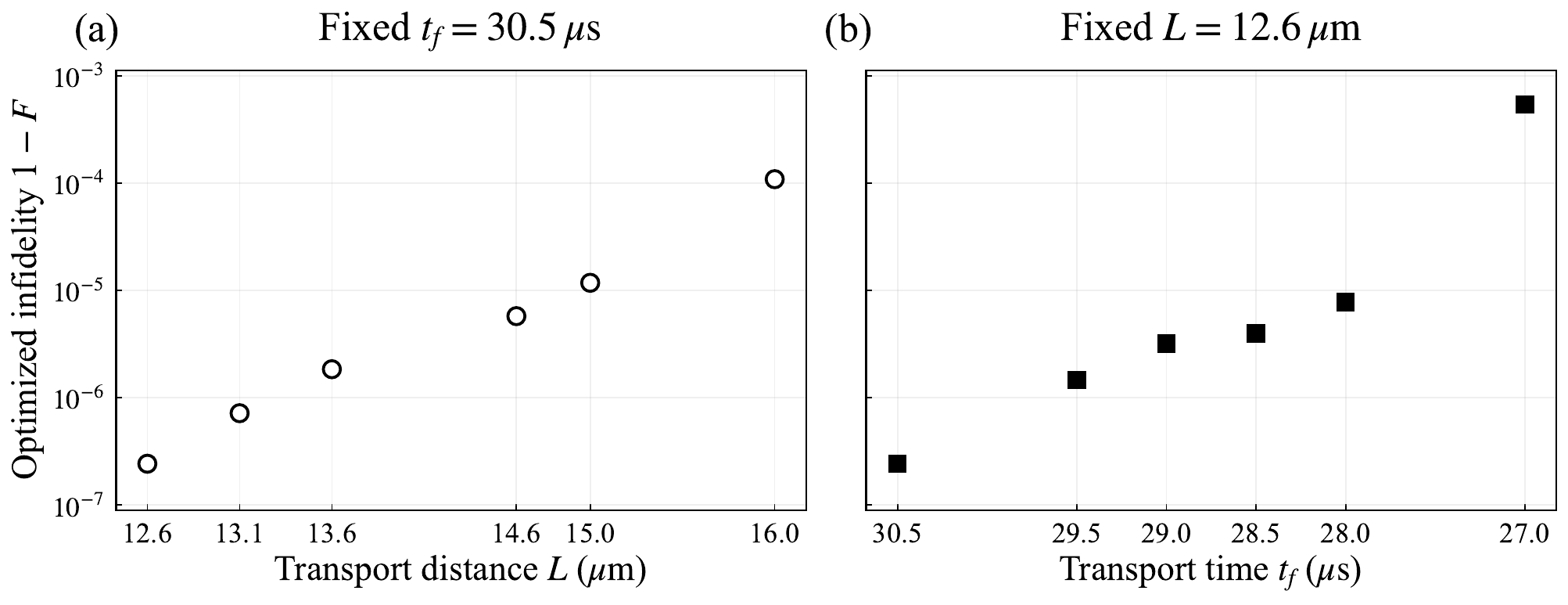}
\caption{Ground-state joint control under varied transport conditions. (a) Distance scan at $t_f=30.5\,\mu$s. (b) Duration scan at $L=12.6\,\mu$m. Each marker denotes a separately optimized waveform followed by independent refined propagation. Neighboring markers are discrete tasks rather than a continuous fidelity frontier.}
\label{fig:scan}
\end{fixedcolumnfigure}
The selected error rises as distance increases or duration decreases across the sampled tasks, although the size of each increment varies. These separately optimized points describe achieved performance, not a continuous fidelity frontier or a fundamental speed limit. The two extreme scan endpoints, $L=16.0\,\mu$m at $30.5\,\mu$s and $L=12.6\,\mu$m at $27.0\,\mu$s, are analyzed through force utilization and centroid deviation in Appendix~\ref{app:force}.

\FloatBarrier
\section{Transport with finite motional populations}
\subsection{Population-weighted same-state objective}
The multistate comparisons retain the baseline task, $L=12.6\,\mu$m and $t_f=30.5\,\mu$s, with the same physical parameters and coefficient bounds as the ground-state study.
To extend the ground-state task to the mixed motional populations relevant after tweezer cooling \cite{Kaufman2012,Spence2022}, an incoherent initial ensemble is described by
\begin{equation}
 \rho_A=\sum_n p_n\ket{\phi_n^A}\bra{\phi_n^A},\qquad
 \nbar=\sum_n np_n.
 \label{eq:rho}
\end{equation}
This is a statistical mixture of initial vibrational states, with no coherence between distinct labels. The temperature parameter $T$ sets its Boltzmann weights independently of the trap-depth energy $U_0$. The weights $p_n$ specify preparation probabilities, whereas $P_n$ in Eq.~\eqref{eq:cumulativeexc} describes the terminal populations for a single propagated input.
The populations are calculated from the static Gaussian eigenenergies,
$p_n\propto\exp[-(E_n-E_0)/(k_BT)]$, and normalized over the retained initial states \cite{Pagano2024}. We characterize the two ensembles by their actual mean occupations, $\nbar\simeq0.16093$ and $1.87316$. The harmonic estimate
$\nbar_{\rm ho}=[\exp(\hbar\omega_0/k_BT)-1]^{-1}$ is useful for scale setting but does not replace the Gaussian-spectrum weights.

A common control is applied to every occupied initial state, as in finite-temperature optimal-control treatments of tweezer transport \cite{Pagano2024}. For the specified state-label mapping, we define the population-weighted same-state fidelity as
\begin{align}
 \F_n&=\left|\bra{\phi_n^B}\Uop(t_f,0)\ket{\phi_n^A}\right|^2,\\
 \F_{\rm ens}&=\sum_n p_n\F_n,\qquad J_T=-\ln\F_{\rm ens}.
 \label{eq:FT}
\end{align}
Its coefficient derivative is
\begin{equation}
 \partial_{\theta_\alpha}J_T
 =-\frac{1}{\F_{\rm ens}}\sum_n p_n\,\partial_{\theta_\alpha}\F_n.
 \label{eq:thermalgrad}
\end{equation}
This target penalizes both loss from the corresponding target state and redistribution between different vibrational labels. It does not constrain the relative phases acquired by different $n$ and consequently does not certify preservation of arbitrary coherent motional superpositions or an internal-state quantum gate. To assess the resulting density operator separately, we also evaluate the squared Uhlmann fidelity \cite{Liang2019},
\begin{equation}
 \F_{\rm U}=\left[\operatorname{Tr}\sqrt{\sqrt{\rho_{\rm tar}}\rho_f\sqrt{\rho_{\rm tar}}}\right]^2,
 \label{eq:uhlmann}
\end{equation}
The final and target density operators are
\begin{align}
 \rho_f&=\sum_n p_n\ket{\psi_n(t_f)}\bra{\psi_n(t_f)},\\
 \rho_{\rm tar}&=\sum_n p_n\ket{\phi_n^B}\bra{\phi_n^B},
\end{align}
with the same normalized thermal weights. Unlike $\F_{\rm ens}$, $\F_{\rm U}$ compares the final and target mixed states without retaining the initial-state labels. Neither measure universally bounds the other. We use $\F_{\rm U}$ only as a post-optimization diagnostic; its matrix evaluation and truncation are specified in Appendix~\ref{app:numerics}.

For example, interchanging two equally populated orthogonal states leaves their mixture unchanged, even though neither input reaches its designated target. This illustrates why the density-operator comparison and the label-preserving objective answer different questions \cite{Liang2019}.

\subsection{Why the initial distribution matters}
Figure~\ref{fig:thermal} compares direct reuse of a ground-state joint control with controls optimized for each ensemble. At $\nbar\simeq0.16$, approximately $86.1\%$ of the population is in $n=0$ and $11.9\%$ in $n=1$. Nevertheless, the ground-state control gives only $\F_{\rm ens}=0.9960$ in the displayed validation, whereas population-weighted optimization gives $0.99993$. The corresponding weighted error decreases from $3.96\times10^{-3}$ to $7.26\times10^{-5}$, a reduction of about $98.2\%$.

For $\nbar\simeq1.87$, the first four weights are approximately $0.349$, $0.227$, $0.148$, and $0.096$. The same ground-state control gives $\F_{\rm ens}=0.814$, while the ensemble-specific control reaches $0.9934$. This decreases the error by about $96.4\%$, to $6.64\times10^{-3}$.

\begin{fixedcolumnfigure}
\includegraphics[width=\columnwidth]{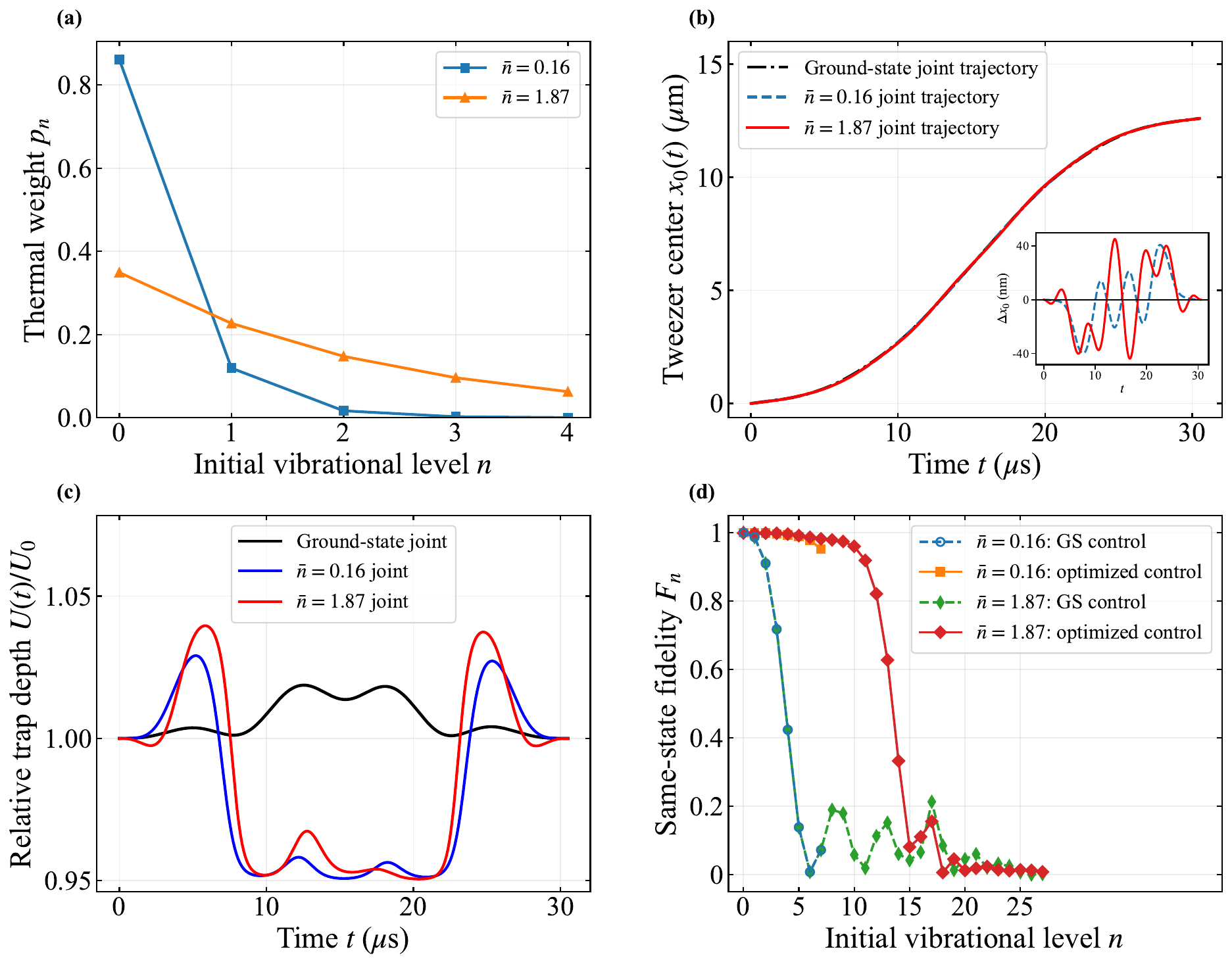}
\caption{Controls and state-resolved transport for two initial motional distributions. (a) Low-lying Gaussian-spectrum populations. (b) Physical center waveforms; the inset shows their differences from the ground-state control. (c) Relative depth waveforms. (d) Same-state probabilities $\F_n$ for reuse of the ground-state control and for separately optimized ensemble controls. Population-weighted fidelities are reported in the text; initial-state truncation checks are given in Appendix~\ref{app:numerics}. For the shared ground-state waveform, overlapping state labels have the same $\F_n$; only their statistical weights differ.}
\label{fig:thermal}
\end{fixedcolumnfigure}

The gain is not uniform over all initial states. In the broader distribution, ensemble optimization extends good same-state transport to roughly $n\lesssim10$, while the higher states remain substantially less accurate. A high weighted score therefore indicates good performance for the significantly occupied states; it does not imply near-unit fidelity for every propagated state.

The center corrections remain small: their maximum deviations from the ground-state joint center are approximately $41$ and $45$ nm. The depth changes more visibly. Ensemble controls approach the lower permitted depth during a substantial central interval and develop higher-depth features on either side. This behavior cautions against interpreting greater confinement as uniformly beneficial. The optimized pulse balances the demands of different initial states through both center motion and time-dependent spectral structure.

\subsection{Stage-resolved weighted errors}
Figure~\ref{fig:weighted} resolves the staged optimization. Its first two panels resolve the contributions
\begin{equation}
 e_n=p_n(1-\F_n),\qquad 1-\F_{\rm ens}=\sum_n e_n
 \label{eq:weightederror}
\end{equation}
when all retained initial states are included. The plotted low-state ranges need not exhaust that sum. The other panels compare the STA, motion, depth-coupled, and final joint stages.

The motion-only stage removes most of the STA error, but the remaining weighted error is reduced further when depth and joint refinement are allowed. In the near-ground ensemble, the full joint result lies below the motion-only result over the displayed contributing states. In the broader ensemble, the state-resolved curves redistribute errors across $n$, while the total weighted error improves. This is a compromise among state-dependent responses rather than independent optimization of each state.

For $\nbar\simeq0.16$, the motion, depth-coupled, and full-joint stages give weighted errors $9.35\times10^{-4}$, $6.28\times10^{-4}$, and $7.26\times10^{-5}$, respectively. For $\nbar\simeq1.87$, the corresponding values are $2.15\times10^{-2}$, $1.45\times10^{-2}$, and $6.64\times10^{-3}$. These reductions show the value of joint refinement when a common waveform must balance several state-dependent nonadiabatic responses. The size of the gain depends on both the initial distribution and the transport task.

Table~\ref{tab:mixedfidelity} compares the label-preserving error with the squared-Uhlmann error for the same frozen controls and validation weights. Both decrease across the three stages. Full joint refinement gives $\F_{\rm U}=0.9999487$ and $0.9979$, alongside $\F_{\rm ens}=0.9999274$ and $0.99336$ for the two distributions. The mixed-state score is higher for these controls, particularly for the broader distribution, but this ordering is result-specific. A high density-operator fidelity alone would not establish accurate transport of each designated vibrational state.

\begin{fixedcolumntable}
\caption{Infidelities of frozen ensemble controls after independent refined propagation. The squared convention is used for $\F_{\rm U}$; the same normalized thermal weights enter both columns.}
\label{tab:mixedfidelity}
\begin{ruledtabular}
\begin{tabular}{llrr}
$\nbar$ & Stage & $1-\F_{\rm ens}$ & $1-\F_{\rm U}$\\
\hline
$\simeq0.16$ & Motion & $9.35\times10^{-4}$ & $8.25\times10^{-4}$\\
 & Depth-coupled & $6.28\times10^{-4}$ & $5.82\times10^{-4}$\\
 & Full joint & $7.26\times10^{-5}$ & $5.13\times10^{-5}$\\
$\simeq1.87$ & Motion & $2.15\times10^{-2}$ & $7.22\times10^{-3}$\\
 & Depth-coupled & $1.45\times10^{-2}$ & $6.59\times10^{-3}$\\
 & Full joint & $6.64\times10^{-3}$ & $2.10\times10^{-3}$
\end{tabular}
\end{ruledtabular}
\end{fixedcolumntable}

\begin{fixedcolumnfigure}
\includegraphics[width=\columnwidth]{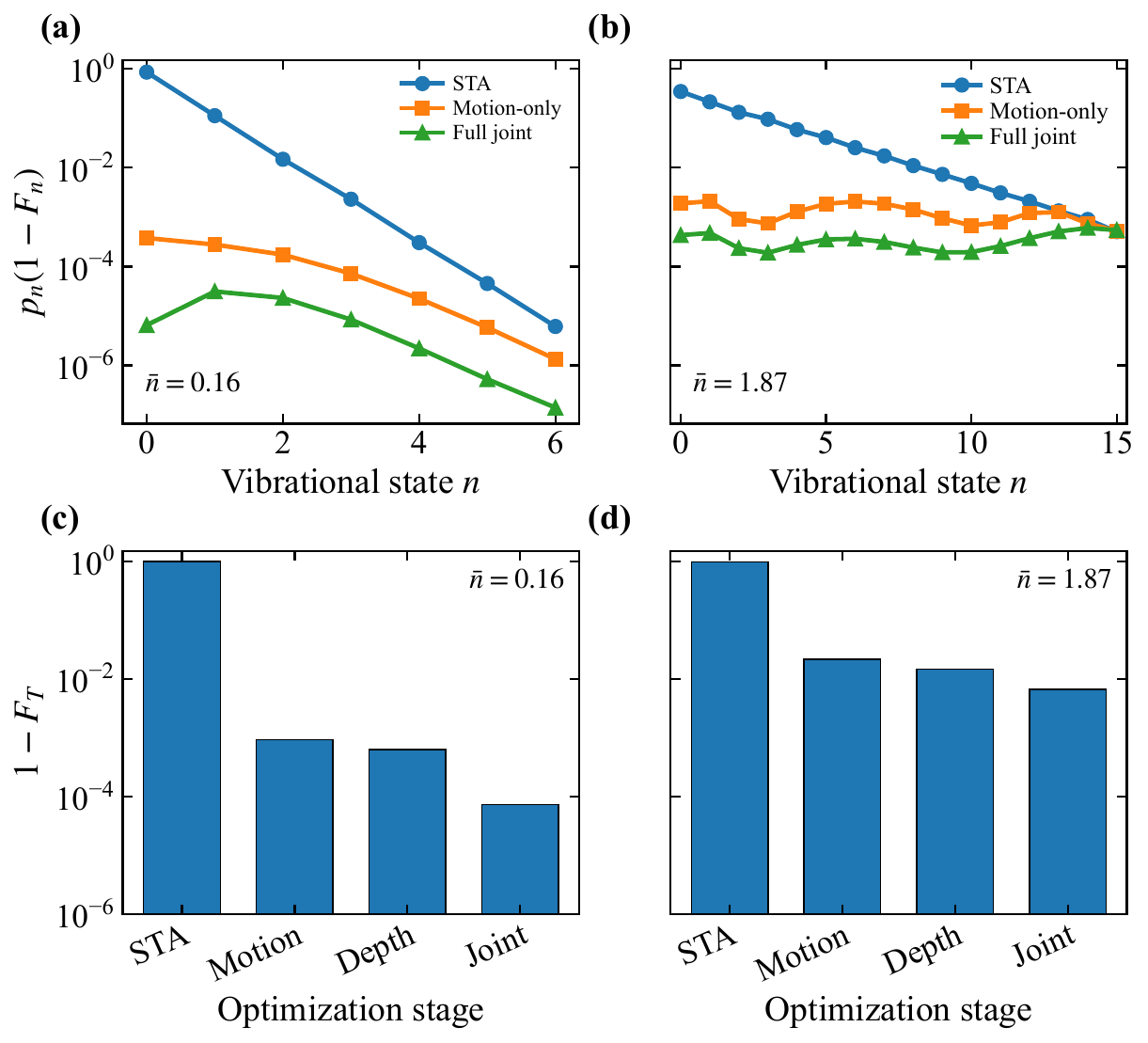}
\caption{Population-weighted error contributions and staged optimization. (a),(b) Contributions $p_n(1-\F_n)$ for the two ensembles, comparing STA, motion-only, and full joint controls. (c),(d) Total weighted infidelity at successive optimization stages. The plotted $F_T$ denotes $\F_{\rm ens}$ in Eq.~\eqref{eq:FT}. The depth stage retains the motion coefficients but reconstructs the physical center using the updated depth.}
\label{fig:weighted}
\end{fixedcolumnfigure}

\FloatBarrier
\section{Robustness to a measurement-informed depth-noise model}
\subsection{Noise model and open-loop implementation}
We next use the $\nbar\simeq0.16$ ensemble to assess depth fluctuations. The task remains $L=12.6\,\mu$m and $t_f=30.5\,\mu$s. The nominal protocol is the noiseless ensemble-optimized joint control; robust refinement uses the same 27-parameter control family with depth-noise scenarios included in its objective. Prior transport-noise analyses distinguish spring-constant or optical-depth fluctuations from position noise and show that their sensitivities can depend on both spectrum and protocol \cite{Lu2018,Lu2020}. In our laboratory's light-intensity-locked, externally controlled operating mode, optical-power measurements identified slow relative drift of order $10^{-3}$ and narrowband components near 2 and 4 MHz. These are characteristic frequencies of the measured power noise under that operating condition, not prescribed trap frequencies. The model is
\begin{align}
 \epsilon_\nu(t)&=\delta_\nu+
 A_2\sin(2\pi f_2t+\varphi_{2,\nu})\nonumber\\
 &\quad+A_4\sin(2\pi f_4t+\varphi_{4,\nu}),\label{eq:noise}\\
 |\delta_\nu|&\le10^{-3},\quad
 A_2=0.0035566,\quad A_4=0.0056368.\nonumber
\end{align}
Here $\nu$ labels a noise scenario specified by $(\delta_\nu,\varphi_{2,\nu},\varphi_{4,\nu})$. The dimensionless quantity $\epsilon_\nu(t)$ is the fractional depth fluctuation, and $\delta_\nu$ is a quasistatic fractional offset held constant during one transport. The phases $\varphi_{2,\nu}$ and $\varphi_{4,\nu}$ are the initial phases in radians of the components at $f_2=2$ MHz and $f_4=4$ MHz. Their dimensionless amplitudes $A_2$ and $A_4$ are relative \emph{peak} modulation amplitudes. The 4 dB difference in measured electrical line powers gives the amplitude ratio $A_4/A_2=10^{4/20}\simeq1.585$ under a common transfer response; the absolute scale is an adopted fractional-depth model value, not an absolute optical-noise calibration. The measurement chain and conversion are detailed in Appendix~\ref{app:robust}.

For every scenario $\nu$, the implemented controls are
\begin{equation}
 U_\nu(t)=U_{\rm des}(t)[1+\epsilon_\nu(t)],
 \qquad x_{0,\nu}(t)=x_{0,\rm des}(t).
 \label{eq:noisycontrol}
\end{equation}
All noise scenarios use the same initial ensemble and target eigenstates of the nominal endpoint traps at depth $U_0$; these states and their thermal weights are not re-prepared for individual realizations.
Here $U_{\rm des}$ and $x_{0,\rm des}$ denote the commanded depth and center, while $U_\nu$ and $x_{0,\nu}$ denote the realized controls in scenario $\nu$. The center is designed once and is not recomputed using the unknown realized depth. During optimization, changes in the \emph{design} depth still modify the design center through Eq.~\eqref{eq:center}. This distinction preserves the open-loop nature of the robustness problem. Here open-loop refers to predetermined transport commands; the laboratory intensity-stabilization feedback remains part of the operating condition used to motivate the noise model. The $5\%$ control bound applies to $U_{\rm des}$; noise is superposed on it.

\subsection{Joint nominal, average, and tail objective}
For each scenario define $\ell_\nu=-\ln\F_{{\rm ens},\nu}$, and let $\ell_{\rm nom}$ denote the corresponding zero-noise loss. Robust optimization minimizes
\begin{equation}
 J_{\rm rob}=0.30\,\ell_{\rm nom}
 +\frac{0.25}{|\mathcal B_{\rm mean}|}
 \sum_{\nu\in\mathcal B_{\rm mean}}\ell_\nu
 +0.45\,\mathcal L_{\rm sw},
 \label{eq:robust}
\end{equation}
where the tail term is a normalized log-sum-exp \cite{Blanchard2021}:
\begin{equation}
 \mathcal L_{\rm sw}
 =\tau\ln\left[\frac{1}{|\mathcal B_{\rm tail}|}
 \sum_{\nu\in\mathcal B_{\rm tail}}e^{\ell_\nu/\tau}\right],
 \quad\tau=10^{-5}.
 \label{eq:softworst}
\end{equation}
The sets $\mathcal B_{\rm mean}$ and $\mathcal B_{\rm tail}$ contain the scenarios used for the mean and soft-worst losses, respectively; $|\mathcal B|$ denotes the number of scenarios in a set. The positive dimensionless smoothing parameter $\tau$ controls the approximation to the maximum scenario loss, recovered as $\tau\to0^+$. The coefficients 0.30, 0.25, and 0.45 are the nominal, mean, and tail weights. The nominal term protects the intended noise-free transfer, the mean term rewards performance across the chosen scenarios, and the soft-worst term gives larger gradient weight to higher-loss scenarios. Here \emph{tail} refers to sampled high-loss cases, rather than a fixed probability quantile of a calibrated noise distribution. The scenario-weighted objective is specific to this work; robust trajectory optimization and disorder-averaged pulse design offer related ways to account for control uncertainty \cite{Propson2022,Araki2023}. The fixed mean set provides a common average objective, while the tail set is enriched between rounds with difficult scenarios. A separate guard set screens candidate updates. The nominal, mean, and tail weights prevent the search from focusing exclusively on one unfavorable realization.

After optimization, both the original nominal control and the selected robust control are propagated on the same 192-scenario grid. It combines $\delta_\nu=-10^{-3},0,10^{-3}$ with $8\times8$ shifted high-frequency phase pairs. These phases differ from the optimization and guard phases. The resulting statistics refer to this finite, deterministically sampled ensemble. Their dependence on the assumed noise model should be kept distinct from claims of robustness to an unknown continuous spectrum \cite{Hocker2014}. Since final acceptance also consults this set, it is an out-of-training validation set with an acceptance role, not an additional untouched statistical test set.

\subsection{Nominal and robust control waveforms}
Figure~\ref{fig:robustcontrols} compares the nominal and robust controls. Their physical centers nearly overlap on the transport-distance scale; the inset resolves nanometer-scale changes. The depth waveforms differ more visibly. Robustness is thus obtained through coordinated changes of the physical center and depth, with the same designed center applied to all noise scenarios.
\begin{fixedcolumnfigure}
\includegraphics[width=\columnwidth]{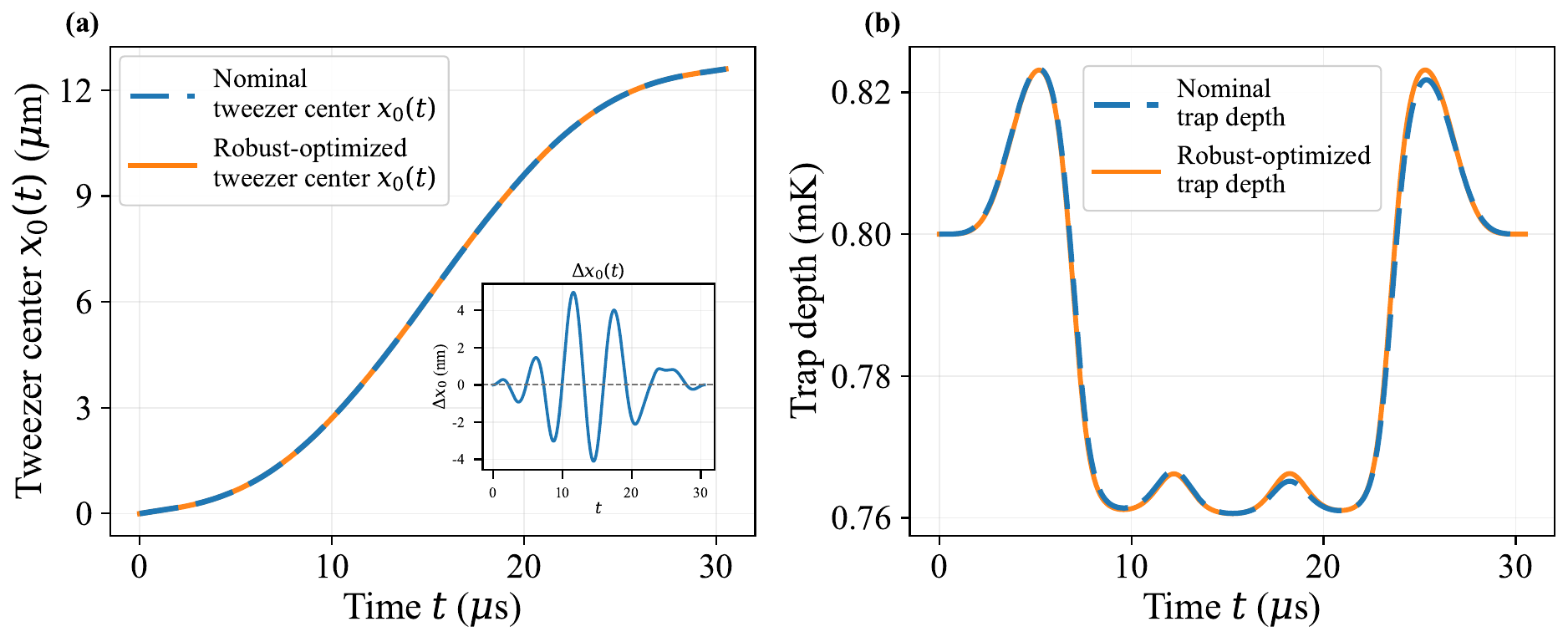}
\caption{Nominal and robust controls for the near-ground ensemble. Blue dashed and orange solid curves denote the nominal and robust protocols. (a) Physical trap center; the inset resolves the center difference $\Delta x_0$. (b) Trap depth expressed as $U/k_B$ in mK. Each pair of waveforms is specified before the noise realization.}
\label{fig:robustcontrols}
\end{fixedcolumnfigure}

\subsection{Paired validation results}
Figure~\ref{fig:robust} shows the noise model, quasistatic response, fidelity distribution, and summary statistics. For the paired set, the mean error is the arithmetic average of $1-\F_{{\rm ens},\nu}$ and the worst sampled error is its largest value. The cumulative distribution at a fidelity threshold gives the fraction of the 192 scenarios whose fidelities do not exceed that threshold. The plotted $\sigma_F$ is the root-mean-square deviation from their mean fidelity. The reported mean error averages infidelities directly, while the mean term in the optimization averages logarithmic losses. The quasistatic response improves most on the unfavorable negative-bias side. The two curves cross on the positive-bias side, so the improvement is not pointwise over every scenario. The cumulative distributions similarly show a substantial improvement of the lower-fidelity tail without establishing stochastic dominance at every quantile.

\begin{fixedcolumnfigure}
\includegraphics[width=\columnwidth]{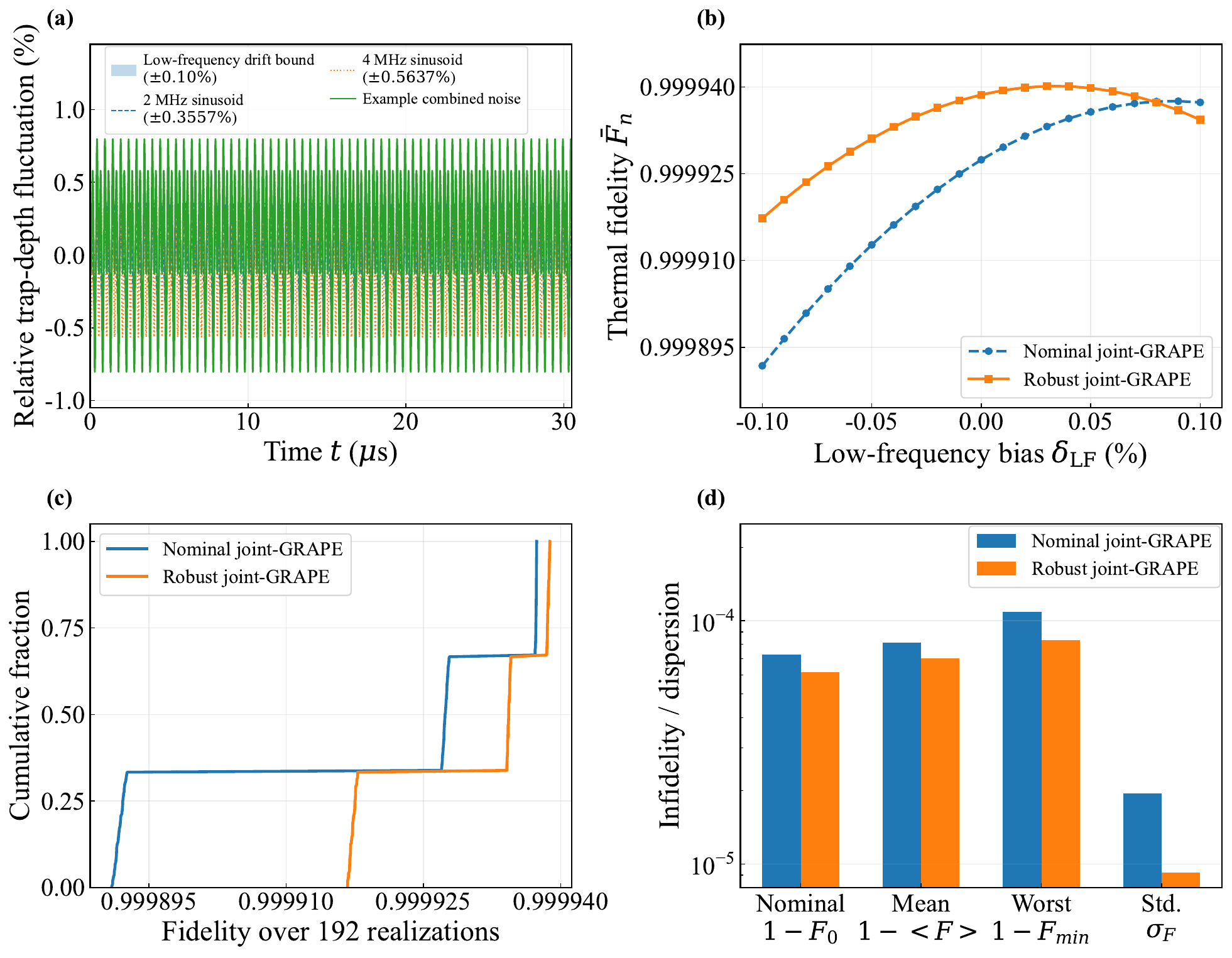}
\caption{Robust transport of the $\nbar\simeq0.16$ ensemble. (a) Adopted depth-noise components and a sample combined waveform. (b) Response to quasistatic bias with both high-frequency components disabled. (c) Cumulative fidelity distributions over 192 paired scenarios, with three low-frequency offsets and $8\times8$ shifted high-frequency phase pairs per offset. (d) Zero-noise error, ensemble mean error, worst sampled error, and fidelity standard deviation. Numerical propagation and sampling settings are given in Appendices~\ref{app:numerics} and~\ref{app:robust}. The panel notation $F_0$ denotes the zero-noise \emph{ensemble} fidelity, not the single-ground-state fidelity of Eq.~\eqref{eq:F0}.}
\label{fig:robust}
\end{fixedcolumnfigure}

As shown in Fig.~\ref{fig:robust}(d), the mean error decreases by approximately $13.9\%$, the worst sampled error by $23.6\%$, and the fidelity standard deviation by $52.8\%$. The larger reduction in tail error and dispersion is consistent with the tail-weighted objective. The zero-noise ensemble error also decreases; the observed robustness gain is not obtained by sacrificing nominal performance in this selected comparison.

The refined paired evaluation gives mean errors of approximately $8.12\times10^{-5}$ and $6.99\times10^{-5}$, and worst sampled errors of $1.09\times10^{-4}$ and $8.33\times10^{-5}$, for nominal and robust controls, respectively. The worst sampled fidelity is not a lower bound over arbitrary noise waveforms or over all continuous phase choices in Eq.~\eqref{eq:noise}.

\FloatBarrier
\section{Discussion and conclusions}
Basis-parameterized GRAPE refines a harmonic STA reference using the full Gaussian wave-function dynamics while keeping the search dimension compact. Seventeen trajectory variables correct the physical center; ten additional bounded depth variables change confinement and the inverse center construction. This separates the numerical resolution needed for propagation from the number of optimized control coefficients.

The ground-state results connect small control corrections to suppression of final motional excitation. Trajectory optimization already performs well in the longer task, while joint center--depth refinement becomes more valuable in the demonstrated faster task. State-resolved final populations and staged search histories support this comparison without assigning the dynamics to a single excited-state channel. The auxiliary trajectory organizes control design but need not coincide with the actual centroid. Instantaneous couplings, spectral phases, and broad intermediate-state participation provide complementary explanations in the appendices.

For initially populated vibrational states, optimizing a common population-weighted objective is substantially more effective than directly reusing a ground-state waveform. The companion squared-Uhlmann fidelity also improves across the optimization stages, showing closer agreement with the target thermal density operator. These fidelity measures retain different information: neither establishes perfect transport of every high-lying state or preservation of arbitrary relative motional phases. Adding a specified depth-noise model further reduces the mean error, sampled tail error, and fidelity dispersion of the near-ground ensemble.

Actuator response and explicit velocity or acceleration constraints would strengthen experimental implementation. Multidimensional propagation would test transverse coupling, as a prior three-dimensional anharmonic-transport study motivates \cite{Hauck2021}, while calibrated optical transfer functions would connect electrical noise observations to quantitative depth fluctuations. Additional basis sizes and independent initializations would extend the present control-space sensitivity checks; enlarged-domain comparisons support the stability of the two extreme tasks' infidelity scales and mean-force diagnostics across the tested domains. A phase-sensitive multistate target would be needed for arbitrary coherent motional superpositions. The transport fidelities are numerical predictions of the stated model, with laboratory measurements supplying the noise-model motivation. Within the one-dimensional model and the specified noise ensemble, coordinated center and confinement control offers a route toward rapid transport with low residual excitation.

Combining the low-excitation motional transport studied here with magic-intensity-based internal-state protection \cite{Yang2016,Sheng2018} and coherent qubit-transfer techniques \cite{Wang2026} could support dynamically connected, zoned neutral-atom processors, provided that the depth modulation remains compatible with qubit-coherence requirements.

\begin{acknowledgments}
This work was supported by the Natural Science Foundation of Hubei Province (Grant No. 2025AFB851), by the National Natural Science Foundation of China (Grant No. 12374330), and by the Fundamental Research Funds for the Central Universities (Grant Nos. 30106260175 and JC2026TS-003).
\end{acknowledgments}

\appendix
\FloatBarrier
\section{Numerical discretization and validation}
\label{app:numerics}
\subsection{Spatial resolution and computational domain}
The kinetic operator is represented spectrally on a periodic Fourier grid \cite{Gharibnejad2020}. For a box of length $L_x$,
\begin{equation}
 \Delta x=\frac{L_x}{N_x},\qquad
 k_{\rm Nyq}=\frac{\pi}{\Delta x},\qquad
 v_{\rm Nyq}=\frac{\hbar k_{\rm Nyq}}{m}.
 \label{eq:nyquist}
\end{equation}
The last quantity is the represented momentum cutoff expressed as a velocity, not a physical speed limit. For the baseline $L_x=20.6\,\mu$m, selected resolutions are listed in Table~\ref{tab:grid}. The harmonic oscillator length is approximately 36 nm and the ground-state root-mean-square width approximately 25 nm. Static spatial resolution alone is insufficient: the minimum-jerk trajectory has peak velocity $15L/(8t_f)\simeq0.775$ m/s, substantially above its mean velocity $L/t_f\simeq0.413$ m/s. The $N_x=4096$ grid cannot represent even this reference peak momentum, although it resolves the localized static ground state.

\begin{fixedcolumntable}
\caption{Spatial resolution and Fourier momentum bandwidth for the baseline box.}
\label{tab:grid}
\begin{ruledtabular}
\begin{tabular}{rrr}
$N_x$ & $\Delta x$ (nm) & $v_{\rm Nyq}$ (m/s)\\
\hline
4096 & 5.029 & 0.456\\
8192 & 2.515 & 0.913\\
12288 & 1.676 & 1.369\\
16384 & 1.257 & 1.826\\
24576 & 0.838 & 2.739\\
32768 & 0.629 & 3.652
\end{tabular}
\end{ruledtabular}
\end{fixedcolumntable}

The baseline domain is $[-4,16.6)\,\mu$m, with $4\,\mu$m guard margins around a $12.6\,\mu$m transfer. The spatial scan fixes this domain and changes only $N_x$. A domain $[-4,L+4)\,\mu$m maintains those margins for other distances. The guard margins separate the transport endpoints from the artificial periodic edges of the Fourier box; enlarging them tests sensitivity to the finite computational region. Fixed-domain grid refinement does not establish convergence with respect to domain size; enlarged-domain comparisons for the two extreme tasks are reported in Appendix~\ref{app:force}.

\subsection{Temporal resolution}
Write $t_{\ell+1/2}=(\ell+1/2)\Delta t$ and $\Delta t=t_f/M$. With $V_\ell=V(t_{\ell+1/2})$, the split step is \cite{Gharibnejad2020}
\begin{equation}
 S_\ell=P_\ell K P_\ell,\quad
 P_\ell=e^{-iV_\ell\Delta t/(2\hbar)},\quad
 K=e^{-ip^2\Delta t/(2m\hbar)}.
 \label{eq:split}
\end{equation}
The kinetic factor is applied by FFT and the potential factors by pointwise multiplication. At $M=6400$, $\Delta t=4.765625$ ns for the baseline duration. The large number of steps per trap period is a useful scale comparison, but fidelity convergence must still be tested directly.

\begin{fixedcolumnfigure}
\includegraphics[width=\columnwidth]{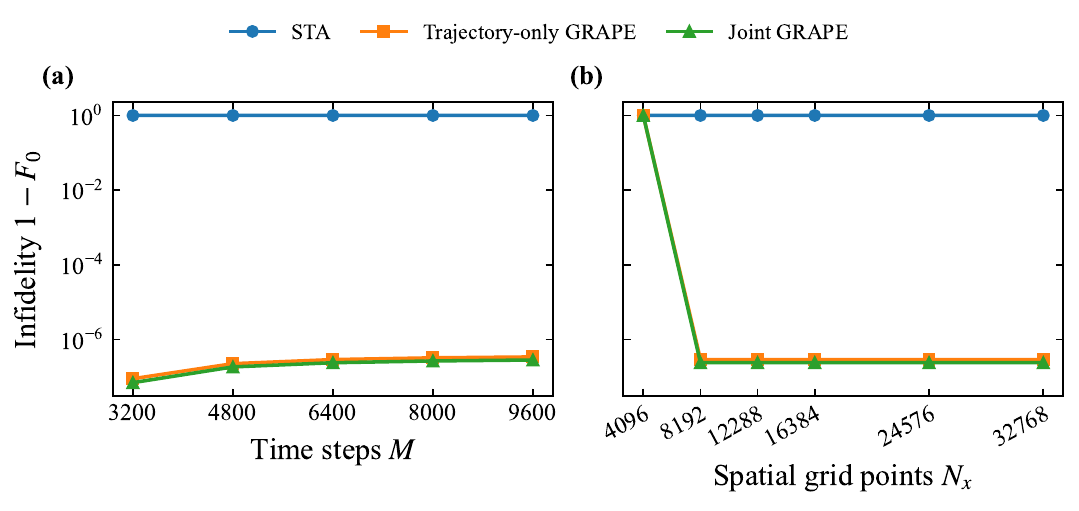}
\caption{Temporal and spatial discretization tests for fixed ground-state controls. Blue circles, orange squares, and green triangles denote STA, trajectory-only GRAPE, and joint GRAPE. (a) Terminal infidelity versus time steps $M$ at $N_x=16384$. The values from $M=6400$ through $9600$ remain on the same error scale and retain the protocol ordering. (b) Terminal infidelity versus spatial points $N_x$ at $M=6400$ in the fixed domain $[-4,16.6)\,\mu$m. No control is reoptimized. The coarsest spatial grid fails because its Fourier momentum bandwidth is inadequate for fast transport.}
\label{fig:time}
\end{fixedcolumnfigure}

Figure~\ref{fig:time}(a) shows that the optimized-control infidelities increase with time resolution before approaching stable values. For the fixed joint waveform, $1-\F_0$ changes from approximately $7.06\times10^{-8}$ at $M=3200$ to $2.41\times10^{-7}$ at $M=6400$. The change is small in absolute fidelity but significant relative to the infidelity. Further refinement from $M=6400$ through $9600$ leaves the optimized errors on the same $10^{-7}$ scale and preserves their ordering. Together with the independent spatial check in panel (b), this supports $M=6400$ as a reliable resolution for the reported baseline error scale and protocol comparison; the last digits are not claimed to be continuum-limit values.

At $N_x=4096$, all three controls give essentially vanishing target overlap, consistent with insufficient Fourier momentum bandwidth and spectral aliasing. At $N_x=8192$, optimized infidelities recover to the $10^{-7}$ scale, and further refinement through $N_x=32768$ gives a stable terminal-fidelity regime containing the adopted $N_x=16384$ grid. These tests concern terminal fidelity; separate checks of time-resolved observables and finite-domain sensitivity for the two extreme tasks are reported in Appendix~\ref{app:force}.

The population-resolved and staged multistate results both use $M=6400$. Their optimized state-preserving fidelities are $0.99993$ and $0.9934$. The recorded high-resolution multistate spatial checks use $N_x=32768$; each waveform is held fixed during validation. The separate $28.5\,\mu$s ground-state comparison in Fig.~\ref{fig:joint} also uses refined repropagation. Noise-ensemble statistics use the common refined setting in Appendix~\ref{app:robust}; they are not inferred from noiseless convergence curves.

\subsection{Initial-state truncation for finite-temperature ensembles}
Let the reference Gaussian-spectrum thermal weights be $p_n$ and
$Q_{N_{\rm mot}}=\sum_{n=0}^{N_{\rm mot}-1}p_n$. The retained weights are
$\widetilde p_n=p_n/Q_{N_{\rm mot}}$. Here $N_{\rm mot}$ counts retained initial states; $Q_{N_{\rm mot}}$ is initial probability coverage, distinct from the terminal populations $P_n$ in Eq.~\eqref{eq:cumulativeexc}. Table~\ref{tab:coverage} gives coverage relative to the archived 16- and 32-state Gaussian-spectrum references. The truncation concerns the initial ensemble; every propagated state still evolves on the full spatial grid. In this subsection only, $p_n$ refers to the finite reference distribution; the main-text populations $p_n$ correspond to the retained, renormalized weights $\widetilde p_n$ defined here.

\begin{fixedcolumntable}
\caption{Initial-state coverage from the Gaussian spectrum. $N_{\rm mot}$ counts retained initial states.}
\label{tab:coverage}
\begin{ruledtabular}
\begin{tabular}{llrr}
$\nbar$ & Stage & $N_{\rm mot}$ & $Q_{N_{\rm mot}}$\\
\hline
$\simeq0.16$ & Optimization & 6 & 0.999992493\\
 & Validation & 8 & 0.999999848\\
$\simeq1.87$ & Optimization & 20 & 0.999781085\\
 & Validation & 28 & 0.999993268
\end{tabular}
\end{ruledtabular}
\end{fixedcolumntable}

Coverage gives a useful bound: if the omitted probability is $\delta=1-Q_{N_{\rm mot}}$ and all omitted-state fidelities lie in $[0,1]$, then
\begingroup
\small
\begin{alignat}{1}
 \F_{{\rm ens},{\rm full}}&=(1-\delta)\F_{{\rm ens},N}
 +\delta\F_{{\rm ens},{\rm tail}},\\
 |\F_{{\rm ens},{\rm full}}-\F_{{\rm ens},N}|&\le\delta.
 \label{eq:truncbound}
\end{alignat}
\endgroup
This bound applies to the stated finite reference distribution; it is not by itself an infinite-dimensional thermal-tail bound and does not control spatial or temporal propagation error. For the near-ground robust comparison, increasing the initial-state count from 8 to 10 and 12 changed checked fidelities by approximately $2.37\times10^{-8}$ and $2.55\times10^{-8}$, respectively.

\subsection{Mixed-state fidelity evaluation}
For the squared Uhlmann fidelity in Eq.~\eqref{eq:uhlmann} \cite{Liang2019}, let $N_{\rm th}$ be the number of retained thermal input states and define the complex overlap matrix $S_{mn}=\bra{\phi_m^B}\psi_n(t_f)\rangle$. With $W=\operatorname{diag}(\widetilde p_0,\ldots,\widetilde p_{N_{\rm th}-1})$ and $S_{\rm th}$ its $N_{\rm th}\times N_{\rm th}$ thermal block, we compute
\begin{equation}
 \F_{\rm U}=\left[\sum_k s_k\!\left(\sqrt W S_{\rm th}\sqrt W\right)\right]^2,
 \label{eq:uhlmannsvd}
\end{equation}
where $s_k$ are singular values. Optimization used $N_{\rm th}=6$ and 20 states for $\nbar\simeq0.16$ and 1.87; independent final evaluation used 8 and 28 states on $N_x=16384$, $M=6400$, with 16 and 32 target eigenstates available for state-resolved checks. Both metrics use the same truncated, renormalized input weights. The target thermal density operator uses the corresponding first $N_{\rm th}$ final eigenstates. No conditional normalization is applied after projection onto that subspace, so leakage lowers $\F_{\rm U}$. The frozen controls were not reoptimized for this supplementary metric.

\FloatBarrier
\section{Discrete adjoint and derivative verification}
\label{app:gradient}
The forward--adjoint construction uses the GRAPE framework \cite{deFouquieres2011}; the expressions below differentiate the stated discrete split-step propagation.
For a single initial state, define
$\ket{\psi_{\ell+1}}=S_\ell\ket{\psi_\ell}$ and
$c=\bra{\phi^B}\psi_M\rangle$. To differentiate $J=-\ln|c|^2$, take
\begin{equation}
 \ket{\lambda_M}=-\frac{c}{|c|^2}\ket{\phi^B},
 \qquad
 \ket{\lambda_\ell}=S_\ell^\dagger\ket{\lambda_{\ell+1}}.
\end{equation}
Then
\begin{equation}
 \partial_u J=
 2\operatorname{Re}\bra{\lambda_{\ell+1}}
 (\partial_u S_\ell)\ket{\psi_\ell}
 \label{eq:discgrad}
\end{equation}
for a control affecting step $\ell$. Because $V_\ell$ is diagonal in position,
\begin{align}
 \partial_u P_\ell
 &=-\frac{i\Delta t}{2\hbar}(\partial_u V_\ell)P_\ell,\\
 \partial_u S_\ell
 &=(\partial_u P_\ell)K P_\ell+P_\ell K(\partial_u P_\ell).
 \label{eq:twokicks}
\end{align}
Both potential half steps must be differentiated. At $\Delta=x-x_0$,
\begin{align}
 \partial_{x_0}V&=-\frac{4U\Delta}{w_0^2}e^{-2\Delta^2/w_0^2},\\
 \partial_UV&=-e^{-2\Delta^2/w_0^2}.
\end{align}
Combining these derivatives with Eqs.~\eqref{eq:chain}--\eqref{eq:jc2} gives the coefficient gradient. For the thermal objective, the terminal adjoint for state $n$ is $-p_nc_n\ket{\phi_n^B}/\F_{\rm ens}$.

An independent check uses the selected full-joint $10+10+7$ coefficient vector and centered finite differences in scaled coordinates. At the audit base point, $\F\simeq0.999929$ and $J\simeq7.10\times10^{-5}$. Let $\bm v$ be a unit direction confined to one coefficient block. We compare
\begin{equation}
\begin{aligned}
 d_{\rm adj}&=\nabla_{\bm z}J\cdot\bm v,\\
 d_{\rm FD}(h)&=\frac{J(\bm z+h\bm v)-J(\bm z-h\bm v)}{2h}.
\end{aligned}
\label{eq:FD}
\end{equation}
The relative directional-derivative error is
\begin{equation}
 E_{\rm rel}=\frac{|d_{\rm adj}-d_{\rm FD}|}{\max(|d_{\rm adj}|,|d_{\rm FD}|,10^{-12})}.
\end{equation}
Figure~\ref{fig:gradcheck} shows the expected second-order decrease for all three blocks. Fitted log--log slopes are 2.0069, 2.0059, and 2.0088 for auxiliary, depth, and residual directions. Table~\ref{tab:gradient} records the derivatives at the smallest tested step. Its purpose is to check agreement between the adjoint and an independent finite-difference calculation for the same direction. The three rows probe distinct coefficient blocks and are not a ranking of their effectiveness. The signs and magnitudes depend on the chosen directions and scaling; the small relative errors verify numerical derivative consistency. The depth direction tests both the explicit potential-depth derivative and its contribution through the reconstructed physical center.
\begin{fixedcolumnfigure}
\includegraphics[width=0.72\columnwidth]{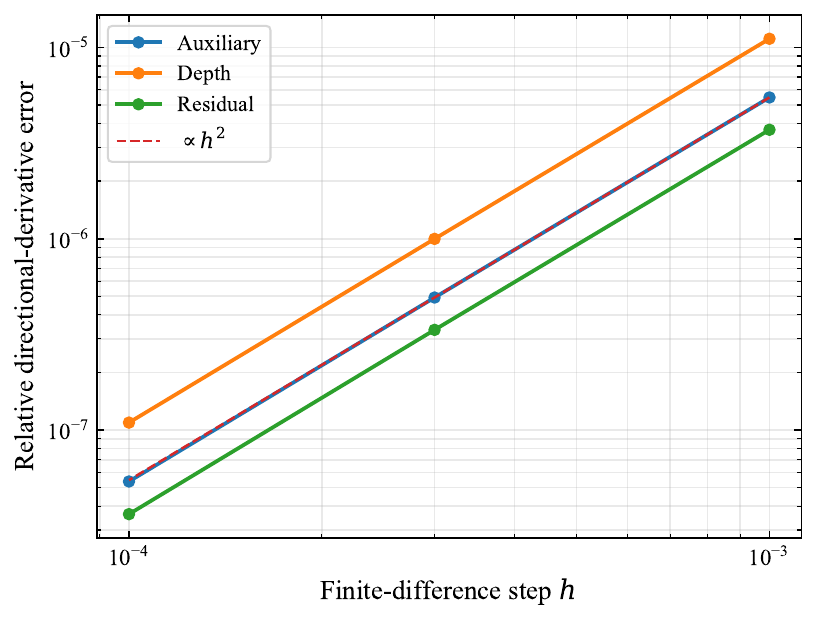}
\caption{Independent finite-difference verification of the discrete-adjoint gradient. Relative directional-derivative error versus centered-difference step $h$ for the auxiliary, depth, and residual coefficient blocks. The dashed reference indicates $h^2$ scaling; the three fitted slopes are approximately 2.}
\label{fig:gradcheck}
\end{fixedcolumnfigure}
\begin{fixedcolumntable}
\caption{Directional derivatives at $h=10^{-4}$. Both derivative columns are expressed in units of $10^{-3}$; errors use the unrounded derivatives. Agreement is assessed within each row, not by comparing derivative magnitudes across blocks.}
\label{tab:gradient}
\begin{ruledtabular}
\small
\setlength{\tabcolsep}{3pt}
\begin{tabular}{lrrr}
Block & $d_{\rm adj}$ & $d_{\rm FD}$ & Relative error\\
\hline
Auxiliary & 4.445383265 & 4.445383025 & $5.39\times10^{-8}$\\
Depth & $-1.641725249$ & $-1.641725070$ & $1.09\times10^{-7}$\\
Residual & $-9.588764434$ & $-9.588764783$ & $3.63\times10^{-8}$
\end{tabular}
\end{ruledtabular}
\end{fixedcolumntable}
This verifies the implemented discrete derivative along three independent block directions. It complements propagation convergence tests and does not itself certify physical fidelity or exhaustively test every coefficient direction.

\FloatBarrier
\section{Nonadiabatic dynamics and local depth ablation}
\label{app:mechanism}
\subsection{Parity structure and instantaneous gaps}
Let $H(t)\ket{n(t)}=E_n(t)\ket{n(t)}$. For nondegenerate instantaneous states in a smooth parallel-transport gauge, $\bra{n}\dot n\rangle=0$, the off-diagonal couplings satisfy \cite{Berry2009}
\begin{equation}
 \kappa_{nm}(t)=\bra{n}\dot m\rangle
 =\frac{\bra{n}\dot H\ket{m}}{E_m-E_n}\quad(n\ne m),
\end{equation}
where $\dot H=\dot x_0\,\partial_{x_0}V+\dot U\,\partial_UV$. The center derivative is odd about the instantaneous center and the depth derivative is even. Direct center-induced couplings from the even ground state connect odd states, whereas depth-induced couplings connect even states. Multistep transfer can nevertheless populate either parity.

After removing instantaneous dynamical phases, the coefficients obey
\begin{equation}
 \dot b_n=-\sum_{m\ne n}b_m\kappa_{nm}
 \exp\!\left[\frac{i}{\hbar}\int_0^t(E_n-E_m)\,\dd t'\right].
 \label{eq:instant}
\end{equation}
Here $b_n$ is the amplitude multiplying the instantaneous state after its dynamical phase has been factored out. A full spectral resolution of a finite-depth potential also includes the continuum; a truncated bound-state expansion is a diagnostic approximation. Coupling magnitude alone cannot determine final fidelity because amplitudes generated at different times accumulate different phases.

Define $\omega_{n0}=(E_n-E_0)/\hbar$, the instantaneous angular transition frequency. Relative to STA, the joint control has shifts
\begin{equation}
 \Delta\omega_{n0}=\omega_{n0}^{J}-\omega_{n0}^{\rm STA},
 \qquad \Delta\Phi_{n0}(t)=\int_0^t\Delta\omega_{n0}(t')\,\dd t'.
\end{equation}
Figure~\ref{fig:mechanism1} shows the coupling magnitudes, transition-frequency shifts, and accumulated phase shifts for the $30.5\,\mu$s, $12.6\,\mu$m ground-state task. The transition-frequency shifts are expressed as $\Delta\omega_{n0}/(2\pi)$ and the accumulated phase shifts as $\Delta\Phi_{n0}/(2\pi)$ cycles. The additional $1/(2\pi)$ scaling of the coupling rates in panel (a) is a plotting convention.
\begin{fixedcolumnfigure}
\includegraphics[width=\columnwidth]{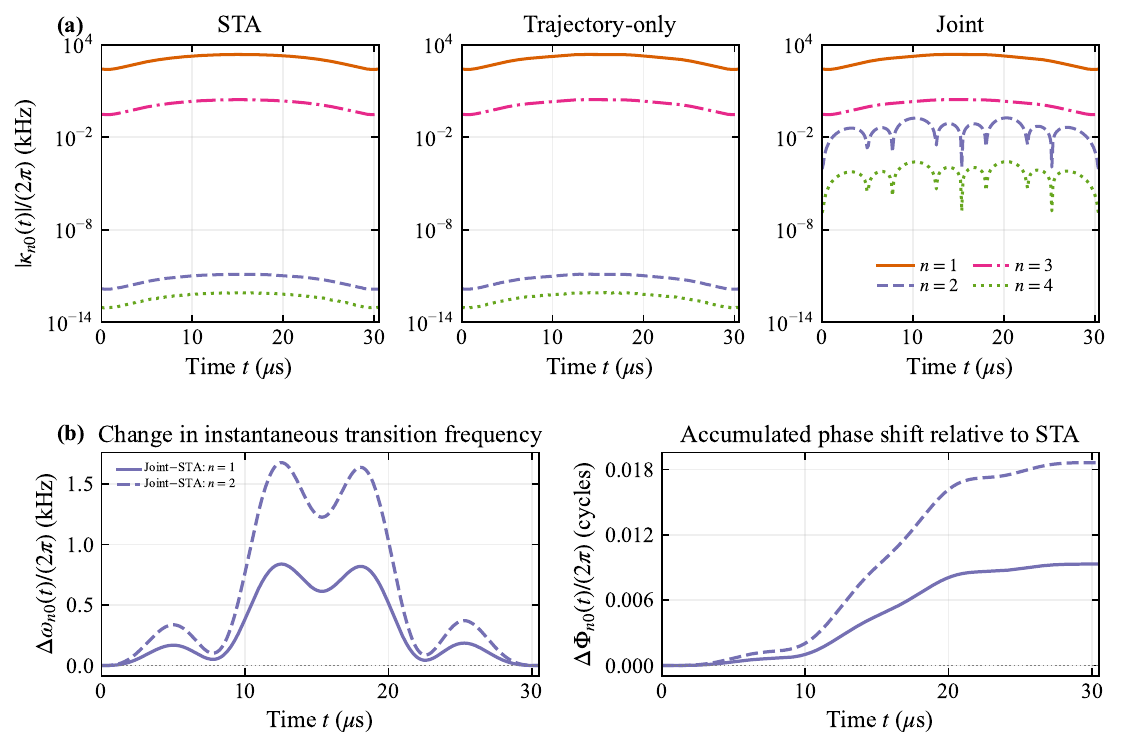}
\caption{Instantaneous-state diagnostics for the longer ground-state task. (a) Coupling magnitudes $|\kappa_{n0}|/(2\pi)$ for the first four excited states under STA, trajectory-only, and joint controls. (b) Joint-control transition-frequency shifts $\Delta\omega_{n0}/(2\pi)$ and accumulated phase shifts $\Delta\Phi_{n0}/(2\pi)$ relative to STA. The displayed low-order channels diagnose control mechanisms rather than close the full multistate dynamics.}
\label{fig:mechanism1}
\end{fixedcolumnfigure}
Pure translation at constant depth does not change the instantaneous spectrum in infinite space. Depth control changes both the gaps and the couplings. The resulting phase shifts describe the instantaneous basis; they do not reconstruct the final error from a few excited states alone.

\subsection{Comoving state populations}
To remove the translational boost, define $\xi=x-x_0(t)$ and use \cite{Torrontegui2011}
\begin{equation}
 \psi_{\rm com}(\xi,t)=
 e^{-im\dot x_0(t)\xi/\hbar}\psi(x_0(t)+\xi,t),
 \label{eq:boost}
\end{equation}
up to a time-dependent global phase. We project this state onto local Gaussian bound states. Interpolation onto a common grid can introduce weak nonorthogonality; the overlap matrix $S=\Phi^\dagger\Phi$ is used for symmetric (L\"owdin) orthogonalization, $\widetilde\Phi=\Phi S^{-1/2}$ \cite{Torun2026}, where columns of $\Phi$ contain the sampled states. All inner products and adjoints in these matrix expressions use the same spatial-quadrature metric; the sampled states are assumed linearly independent, so that $S$ is positive definite. The cumulative captured population is
\begin{equation}
 P_{<K}^{\rm com}(t)=
 \sum_{n=0}^{K-1}|\avg{\widetilde\phi_n|\psi_{\rm com}(t)}|^2.
\end{equation}
Here $K$ counts included states, including $n=0$. The complement $1-P_{<K}^{\rm com}$ is probability outside the displayed subspace. It may occupy higher bound states or unbound components, so a small low-state population is not by itself evidence that the atom has escaped. This time-dependent, boosted-basis quantity differs from the terminal laboratory-frame population $P_n$ in Eq.~\eqref{eq:cumulativeexc}. Figure~\ref{fig:mechanism2}(a) shows these comoving populations; its second panel is defined next.

\subsection{Time-local intervention}
For a window-center time $\tau$ and full temporal width $W>0$, let $u=(t-\tau)/W$ and define
\begin{equation}
 w_{\tau,W}(t)=
 \begin{cases}
 \tfrac12[1+\cos(2\pi u)],& |u|\le \tfrac12,\\
 0,& |u|>\tfrac12 .
 \end{cases}
\end{equation}
The window-center time $\tau$ is distinct from the dimensionless smoothing parameter in Eq.~\eqref{eq:softworst}.
We define a local depth intervention using this continuously differentiable raised-cosine window to remove the optimized depth increment:
\begin{equation}
 U_{\rm ab}(t)=U_0+[1-w_{\tau,W}(t)][U_J(t)-U_0].
\end{equation}
Here $U_0$ is the fixed reference depth and $U_J(t)$ is the selected joint-control depth. Since $0\le w_{\tau,W}\le1$, $U_{\rm ab}$ interpolates between these depths; at the window center it returns to $U_0$ without switching off the trap. The selected physical center $x_J(t)$ is held fixed, without inverse reconstruction or reoptimization, making this diagnostic distinct from the depth-coupled optimization stage. Independent propagation using the same initial and target states gives $\varepsilon_{\rm ab}(\tau;W)=1-\F_{\rm ab}$, compared with the matched joint error $\varepsilon_J=1-\F_J$ through
\begin{equation}
 \Lambda_U=\log_{10}\frac{\varepsilon_{\rm ab}}{\varepsilon_J},
 \qquad \Delta\varepsilon_U=\varepsilon_{\rm ab}-\varepsilon_J.
\end{equation}
For positive terminal infidelities, positive $\Lambda_U$ denotes increased terminal error, $\Lambda_U=1$ denotes a tenfold increase, and negative values indicate improvement after intervention. Figure~\ref{fig:mechanism2}(b) compares this intervention penalty with depth amplitude and checks its time structure for three window widths. The width-specific normalization in Fig.~\ref{fig:mechanism2}(b) compares timing and shape, rather than the absolute penalty between window widths.
\begin{fixedcolumnfigure}
\includegraphics[width=\columnwidth]{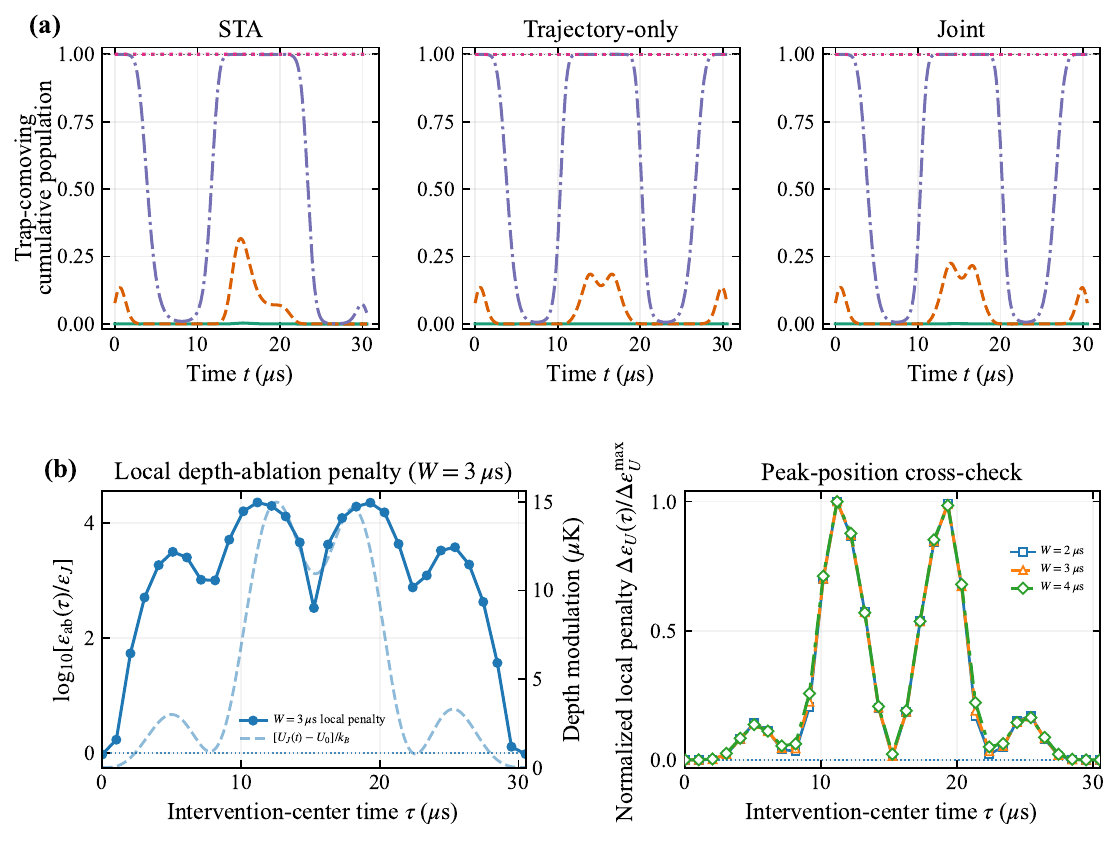}
\caption{Comoving populations and supplementary local depth intervention for $L=12.6\,\mu$m, $t_f=30.5\,\mu$s. (a) STA, trajectory-only, and joint control, from left to right. Green solid curves show the comoving ground-state population; orange dashed, purple dash-dotted, and magenta dotted curves show $P_{<K}^{\rm com}$ for $K=5,20,90$. (b) Left: blue circles show the terminal-error penalty $\Lambda_U(\tau)$ for $W=3\,\mu$s, while the light-blue dashed curve shows the optimized increment $[U_J(t)-U_0]/k_B$ in $\mu$K on the right axis. Right: $\Delta\varepsilon_U(\tau)/\max_\tau\Delta\varepsilon_U(\tau)$ for $W=2,3,4\,\mu$s, normalized separately for each width. Every intervention holds the physical joint center fixed.}
\label{fig:mechanism2}
\end{fixedcolumnfigure}
The minimum population captured by 20 states is approximately $0.00125$, $0.00660$, and $0.00678$ for STA, trajectory-only, and joint controls, respectively; the first 90 states capture more than $0.99996$ throughout the inspected evolution. Intermediate dynamics therefore involve many states and cannot be read as weak excitation confined to a few low levels.

These populations depend on the comoving frame. The transport objective remains the laboratory-frame overlap with the final static target. Equation~\eqref{eq:center} does not impose $\dot x_0(t_f)=0$, so the terminal boosted-basis ground-state population need not equal $\F_0$. Broad intermediate populations are compatible with high final target fidelity.

The strongest local penalties occur near $11$--$12\,\mu$s and $19$--$20\,\mu$s, with similar main locations for $W=2,3,4\,\mu$s. The dashed depth curve measures modulation amplitude; the blue penalty measures sensitivity of the final outcome to a local intervention. Their peaks need not coincide. The intervention changes subsequent coherent evolution and is not an instantaneous transition probability. It combines depth amplitude, evolving state composition, couplings, and phase accumulation, and serves as a supplementary diagnostic rather than the primary evidence for joint-control performance.

\FloatBarrier
\section{Force bound, curvature, and reference accessibility}
\label{app:force}
This appendix separates the acceleration demand of the auxiliary trajectory from the force acting on the propagated wave packet. The resulting ratios help interpret the demanding distance--duration combinations, while the terminal state overlap remains the criterion for motional-state-preserving transport.
\subsection{Gaussian force and reference acceleration}
For $\Delta=x-x_0(t)$, the restoring force and curvature are
\begin{align}
 F(\Delta,t)&=-\frac{4U(t)\Delta}{w_0^2}e^{-2\Delta^2/w_0^2},\\
 V''(\Delta,t)&=\frac{4U(t)}{w_0^2}
 \left(1-\frac{4\Delta^2}{w_0^2}\right)e^{-2\Delta^2/w_0^2}.
\end{align}
The maximum force magnitude and acceleration scale are
\begin{equation}
 F_{\max}[U]=\frac{2U}{w_0}e^{-1/2},
 \qquad a_{\max}[U]=\frac{F_{\max}[U]}{m}.
\end{equation}
The maximum occurs at $|\Delta|=w_0/2$, where the local curvature vanishes, as shown in Fig.~\ref{fig:curvature}. This softens a force-balanced equilibrium but does not determine the quantum transport fidelity.

Differentiating Eq.~\eqref{eq:qref} gives $\ddot q_{\rm ref}=60Ls(1-s)(1-2s)/t_f^2$. Its extrema occur at $s=(3\pm\sqrt3)/6$. The fixed-depth minimum-jerk reference load is therefore
\begin{equation}
 \eta_{\rm MJ}^{(0)}=\frac{\max_t|\ddot q_{\rm ref}|}{a_{\max}[U_0]}
 =\frac{10L}{\sqrt3\,a_{\max}[U_0]t_f^2}.
 \label{eq:eta}
\end{equation}
If the actual centroid were required to follow this reference exactly, $\eta_{\rm MJ}^{(0)}\le1$ would be necessary. It is not sufficient, because the evolving wave packet must produce the required average force. Increasing the scan distance from $12.6$ to $16.0\,\mu$m at $30.5\,\mu$s increases this reference load from about $0.824$ to $1.046$; shortening to $27.0\,\mu$s at $12.6\,\mu$m gives about $1.052$. These are minimum-jerk reference loads, not the optimized-auxiliary ratios below.

For an optimized auxiliary coordinate and depth, define
\begin{align}
 \eta_{q_c}^{(0)}(t)&=\frac{|\ddot q_c(t)|}{a_{\max}[U_0]},\\
 \eta_{q_c}^{(U)}(t)&=\frac{|\ddot q_c(t)|}{a_{\max}[U(t)]},\\
 \eta_{\rm com}(t)&=\frac{|\avg{F(t)}|}{F_{\max}[U(t)]}.
\end{align}
The first two quantities measure auxiliary acceleration relative to constant- and time-dependent force scales. The last uses the actual propagated wave function. Ehrenfest's relation \cite{Bondar2012} gives
\begin{align}
 a_{\rm com}(t)&=\frac{\dd^2\avg{x}}{\dd t^2}=\frac{\avg{F(t)}}{m},\\
 \avg{F(t)}&=\int |\psi(x,t)|^2F[x-x_0(t),t]\,\dd x.
\end{align}
In general $\avg{F}\ne F(\avg{x}-x_0)$. For a normalized state, $|\avg F|\le\avg{|F|}\le F_{\max}$, hence $\eta_{\rm com}\le1$. Figure~\ref{fig:forceuse} compares these ratios for the two extreme scan endpoints.

The first two ratios obey the pointwise identity $\eta_{q_c}^{(0)}=[U(t)/U_0]\eta_{q_c}^{(U)}$, but their separate maxima need not occur at the same time. The actual-force ratio uses the normalized propagated probability density and differs because $q_c$ is not imposed on the Gaussian dynamics. The sampled maxima for the two tasks are listed in Table~\ref{tab:forcepeaks}; the 16.0-$\mu$m protocol has $\eta_{q_c}^{(U)}<1$ at every recorded time, whereas the 27.0-$\mu$s auxiliary demand exceeds the instantaneous force scale briefly. Neither statement alone determines final fidelity. The actual mean-force ratio remains below unity, as required by the pointwise force bound.

\begin{fixedcolumnfigure}
\includegraphics[width=\columnwidth]{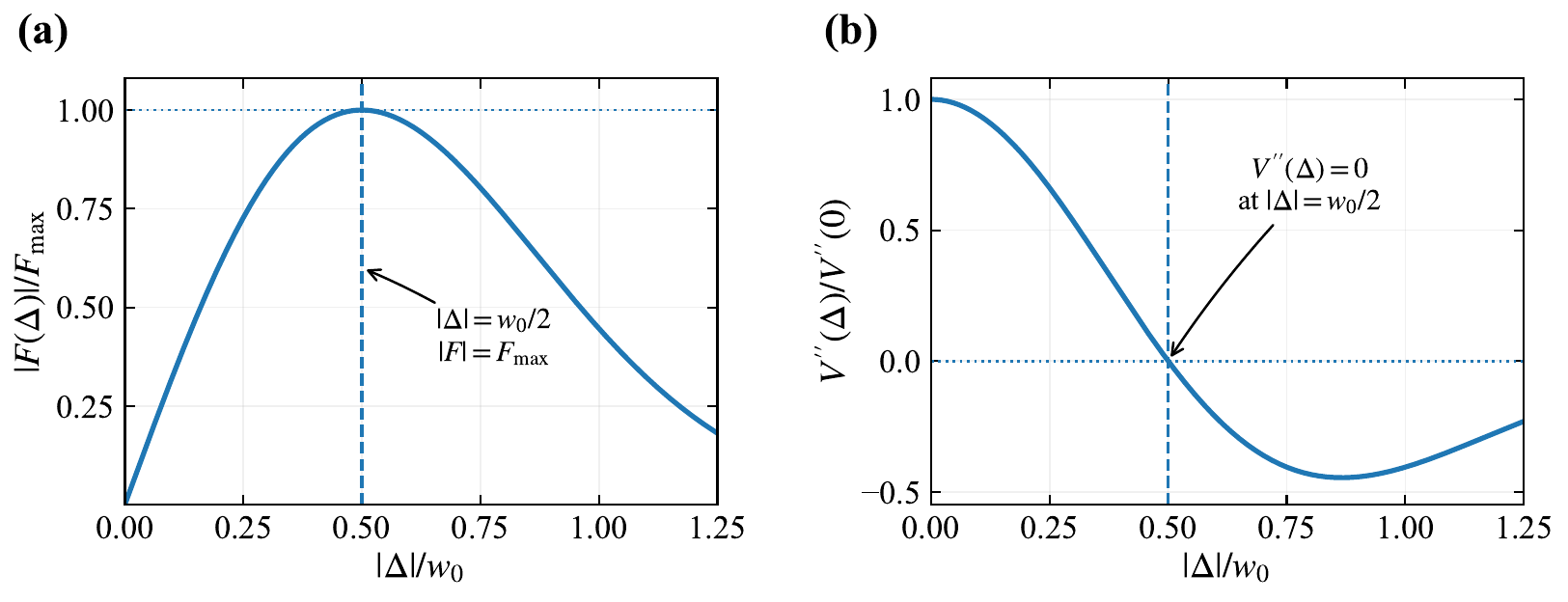}
\caption{Gaussian restoring-force magnitude and local curvature versus displacement from the trap center. The maximum force occurs at $|\Delta|=w_0/2$, where the curvature changes sign.}
\label{fig:curvature}
\end{fixedcolumnfigure}

\begin{fixedcolumntable}
\caption{Sampled force-use maxima and terminal ground-state errors at $M=6400$. Each maximum is taken independently over 6401 recorded times. The tasks use separate final joint controls.}
\label{tab:forcepeaks}
\begin{ruledtabular}
\begin{tabular}{lrr}
Quantity & $16.0\,\mu$m, $30.5\,\mu$s & $12.6\,\mu$m, $27.0\,\mu$s\\
\hline
$\max\eta_{q_c}^{(0)}$ & 1.0372 & 1.0729\\
$\max\eta_{q_c}^{(U)}$ & 0.9905 & 1.0268\\
$\max\eta_{\rm com}$ & 0.9654 & 0.9667\\
$U/U_0$ range & 0.9771--1.0478 & 0.9564--1.0472\\
$1-\F_0$ & $1.086\times10^{-4}$ & $5.379\times10^{-4}$
\end{tabular}
\end{ruledtabular}
\end{fixedcolumntable}

\subsection{Dynamical deviation from the reference trajectory}
Let $d(t)=\avg{x(t)}-q_c(t)$. To isolate the deviation from a locally linear force, define the dimensional anharmonic acceleration remainder
\begin{equation}
 r_{\rm anh}(t)=\frac{\avg{F(t)}}{m}
 +\omega^2(t)[\avg{x(t)}-x_0(t)].
\end{equation}
Substituting Eq.~\eqref{eq:center} into Ehrenfest's relation yields the exact centroid identity
\begin{equation}
 \ddot d+\omega^2(t)d
 =\omega^2(t)\delta x_{\rm res}(t)+r_{\rm anh}(t).
 \label{eq:forcedcentroid}
\end{equation}
This has the structure of a driven oscillator with time-dependent frequency, but $r_{\rm anh}$ depends on the evolving quantum state. It is not a closed harmonic approximation. In the ideal harmonic case with no residual correction, $r_{\rm anh}=0$ and zero initial $d,\dot d$ give $d=0$.

\begin{figure}[t]
\includegraphics[width=\columnwidth]{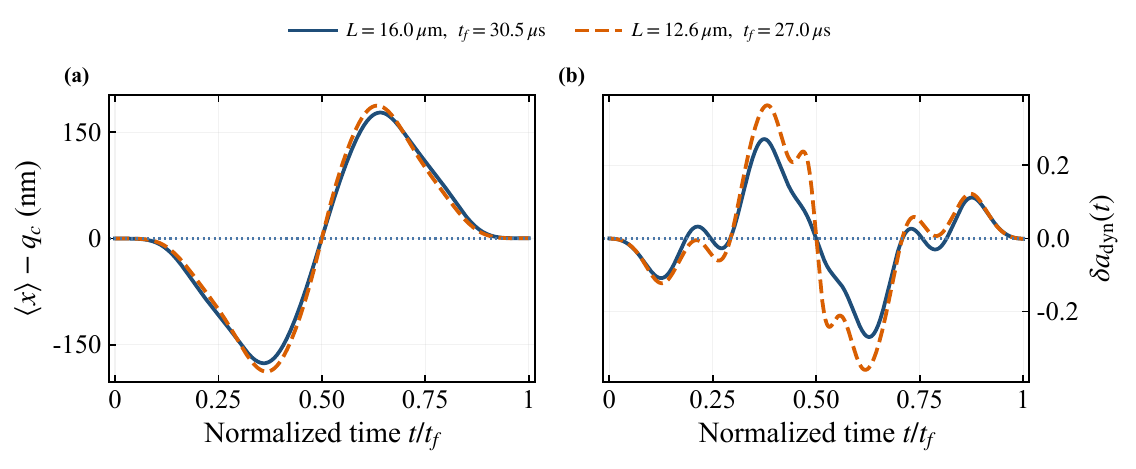}
\caption{Deviation of the actual Gaussian dynamics from the auxiliary coordinate for the two extreme tasks in Fig.~\ref{fig:forceuse}. (a) Centroid displacement $d=\avg{x}-q_c$. (b) Signed acceleration deviation $\delta a_{\rm dyn}$ defined in Eq.~\eqref{eq:deltadyn}. The horizontal coordinate is normalized time $t/t_f$.}
\label{fig:centroiddelta}
\end{figure}

The plotted dimensionless acceleration deviation is
\begin{equation}
 \delta a_{\rm dyn}(t)=
 \frac{a_{\rm com}(t)-\ddot q_c(t)}{a_{\max}[U(t)]}
 =\frac{\ddot d(t)}{a_{\max}[U(t)]}.
 \label{eq:deltadyn}
\end{equation}
It is a signed difference, unlike the absolute force ratios. It includes both the residual center correction and departure from the linear-force response and must not be identified with $r_{\rm anh}/a_{\max}$ alone. Figure~\ref{fig:centroiddelta} displays both the position and acceleration deviations.

The position deviation reaches roughly $180$ nm in both tasks, while the dimensionless acceleration deviation has appreciable time structure. Thus the auxiliary coordinate remains useful for designing controls without being the actual atomic trajectory. Endpoint transport and high final fidelity are compatible with finite intermediate deviations from this reference.

\subsection{Numerical scope of the extreme-task diagnostics}
Both extreme-task diagnostics use selected 27-parameter joint controls. The main validation uses $M=6400$ and $\Delta x=1.26$ nm. The 16.0-$\mu$m task uses $N_x=19090$ and $\Delta t=4.77$ ns; the 27.0-$\mu$s task uses $N_x=16384$ and $\Delta t=4.22$ ns. Independent repropagation reproduces the terminal states and centroid curves of the selected fixed controls.

Holding each selected control and spatial grid fixed, temporal refinement changes the terminal infidelities as shown in Table~\ref{tab:extremetime}. Between $M=6400$ and 25600, the 16.0-$\mu$m error changes by about $5.82\%$, while the 27.0-$\mu$s error changes by about $0.67\%$. The later increments diminish in both cases. Thus $M=6400$ captures the error scale and the qualitative comparison, but its last digits are not continuum-time estimates, especially for the longer-distance task. Table~\ref{tab:extremetime} concerns terminal infidelity; the temporal sensitivity of the dynamical curves is assessed separately below.

\begin{fixedcolumntable}
\caption{Temporal refinement of the extreme-task terminal infidelity with each spatial grid and control held fixed.}
\label{tab:extremetime}
\begin{ruledtabular}
\begin{tabular}{rrr}
$M$ & $16.0\,\mu$m, $30.5\,\mu$s & $12.6\,\mu$m, $27.0\,\mu$s\\
\hline
6400 & $1.09\times10^{-4}$ & $5.38\times10^{-4}$\\
12800 & $1.14\times10^{-4}$ & $5.41\times10^{-4}$\\
25600 & $1.15\times10^{-4}$ & $5.42\times10^{-4}$
\end{tabular}
\end{ruledtabular}
\end{fixedcolumntable}

Finite-domain sensitivity was assessed for the two extreme tasks, $(L,t_f)=(16.0\,\mu\mathrm{m},30.5\,\mu\mathrm{s})$ and $(12.6\,\mu\mathrm{m},27.0\,\mu\mathrm{s})$, by increasing the guard margins beyond both transport endpoints from approximately $4$ to $6$, $8$, and $10\,\mu$m. Each selected control and the spatial step were held fixed. The full domain scan used $M=6400$, and the comparison between the original and largest domains was repeated at $M=12800$. In these comparisons, the observed absolute differences in terminal infidelity were below $10^{-14}$, while the maximum sampled differences in the mean-force utilization $\eta_{\rm com}(t)$ were below $10^{-13}$. Repreparing the initial and target ground states in the largest domain produced negligible changes in these results. Thus, the reported infidelity scales and mean-force diagnostics are stable across the tested domains.

The centroid exhibits a small but distinguishable domain dependence. Increasing the margins from $4$ to $10\,\mu$m changes the centroid by at most approximately $0.13$ nm and $1.11\times10^{-4}$ nm for the longer-distance and shorter-duration tasks, respectively. At $M=6400$, the remaining centroid difference between the $8$- and $10\,\mu$m margins is approximately $2.49\times10^{-4}$ nm for the longer-distance task. These changes are much smaller than the approximately $180$ nm intermediate deviations from the auxiliary trajectory and preserve their qualitative interpretation.

Temporal sensitivity was evaluated separately by comparing the curves at common physical times. Doubling $M$ from 6400 to 12800 gives maximum sampled centroid changes of approximately $0.040$ and $0.023$ nm, and maximum changes in $\eta_{\rm com}(t)$ of approximately $9.35\times10^{-5}$ and $3.59\times10^{-5}$ for the longer-distance and shorter-duration tasks, respectively. These differences remain essentially unchanged after domain enlargement. The comparisons support the qualitative dynamical analysis at $M=6400$; the precision of the terminal infidelities remains subject to the temporal-refinement results in Table~\ref{tab:extremetime}.

\begin{fixedcolumnfigure}
\includegraphics[width=\columnwidth]{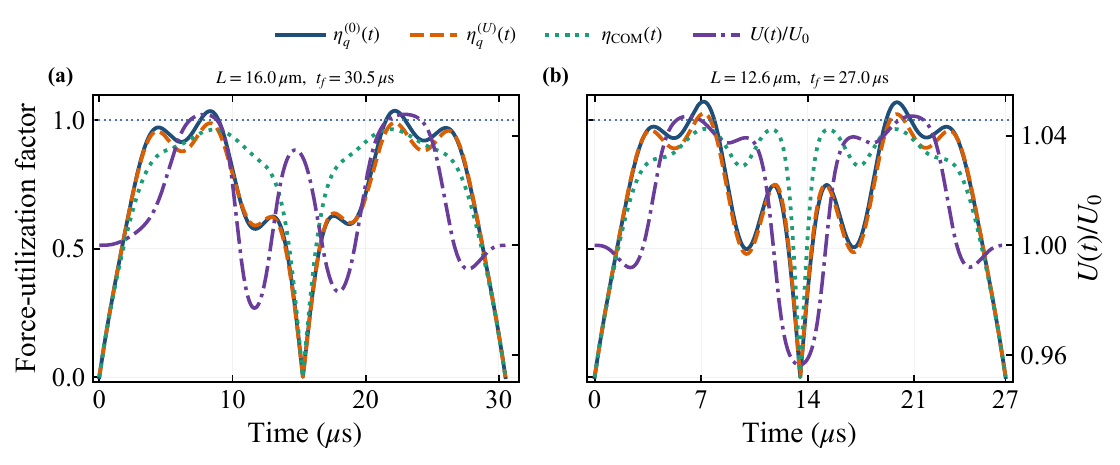}
\caption{Force utilization for the two extreme jointly optimized tasks: (a) $L=16.0\,\mu$m, $t_f=30.5\,\mu$s; (b) $L=12.6\,\mu$m, $t_f=27.0\,\mu$s. The legend's $\eta_q^{(0)}$ and $\eta_q^{(U)}$ denote $\eta_{q_c}^{(0)}$ and $\eta_{q_c}^{(U)}$, respectively; $\eta_{\rm COM}$ denotes $\eta_{\rm com}$. The right axes show $U/U_0$. The horizontal unit line marks equality between an auxiliary-acceleration demand and a force scale; it is not a target-state fidelity threshold.}
\label{fig:forceuse}
\end{fixedcolumnfigure}

\subsection{Conditional kinematic bounds}
A rest-to-rest problem with constant acceleration bound $A$ has maximum distance $At_f^2/4$, obtained with opposite accelerations in the two halves \cite{Chen2011}. The minimum-jerk reference threshold gives $\sqrt3 At_f^2/10$, smaller by a factor $5/(2\sqrt3)\simeq1.443$. For a prescribed time-dependent bound $A(t)=a_{\max}[U(t)]$, the switching time $t_s$ obeys
\begin{equation}
 \int_0^{t_s}A(t)\,\dd t=\int_{t_s}^{t_f}A(t)\,\dd t,
\end{equation}
and the conditional kinematic upper bound is
\begin{equation}
 L_{\rm kin}[U]=\int_0^{t_f}|t-t_s|A(t)\,\dd t.
\end{equation}
This assumes bounded acceleration and zero endpoint velocities. It does not ensure realization by a Gaussian wave packet or return to a specified motional state, and it does not optimize over all allowed depth waveforms.

\section{Noise sampling and robust-control protocol}
\label{app:robust}
\subsection{Measurement scope}
The optical-power measurement was made with the laboratory light-intensity lock engaged and in the externally controlled operating mode used to motivate the transport-noise model. In this arrangement, an LB1005 feedback controller drives the acousto-optic modulator (AOM) to stabilize the light intensity. A Thorlabs PDA36A2 photodetector at the 0 dB electronic-gain setting monitors the output. Its time-domain signal, observed through an NI acquisition system, suggests slow relative fluctuations of order $10^{-3}$. With a DC block inserted before the spectrum analyzer, the same detector signal shows narrow lines near 2 and 4 MHz, at electrical readings near $-65$ and $-61$ dBm. These frequencies are features measured in the locked operating condition. The detector response, DC-block transfer, and losses were not fully calibrated, so the electrical readings alone do not determine absolute optical fractional-noise amplitudes.

The relative high-frequency amplitudes can nevertheless be related to the measured line-power difference. For a sinusoidal voltage component measured with the same impedance and transfer response at both frequencies, electrical power in dBm obeys $P_j\propto V_{j,\rm rms}^2$; peak voltage is $\sqrt2 V_{j,\rm rms}$. Thus the 4 dB difference gives $V_{4,\rm pk}/V_{2,\rm pk}=10^{4/20}\simeq1.585$, and the adopted model values satisfy $A_4/A_2\simeq1.585$. Converting either electrical peak voltage into an \emph{absolute} fractional optical-power amplitude would also require the detector DC reference and the frequency-dependent response of the complete measurement chain. We therefore select $A_2=0.0035566$ as a model fractional peak amplitude and set $A_4=0.0056368$ using the observed ratio; these values are not an absolute calibration inferred solely from $-65$ and $-61$ dBm. Since tweezer depth is proportional to optical power in this model, the selected fractional power amplitudes are used as fractional depth amplitudes. The 0 dB detector gain setting is not a 0 dBm optical power.

The same robust-control construction can be applied to a different measured spectrum by replacing the frequencies, amplitudes, or phase distribution in $\epsilon_\nu(t)$ and rebuilding the scenario ensembles. That would require new optimization and validation; the present numerical robustness claims apply to the specified two-line and quasistatic model. Broadband noise, beam pointing, and waist fluctuations are not included here.

\subsection{Mean, tail, guard, and final ensembles}
The mean scenarios define the average loss, tail scenarios emphasize difficult realizations, guard scenarios screen candidate updates, and the final ensemble compares the selected controls under the same perturbations.
The fixed mean set contains five offsets,
$\delta=(-1,-0.5,0,0.5,1)\times10^{-3}$,
and two phase pairs, $(0,0)$ and $(\pi,\pi)$, giving 10 scenarios. The initial tail set contains the same scenarios. Between accepted rounds, an 80-scenario search combines those five offsets with independent phases from $\{0,\pi/2,\pi,3\pi/2\}$. A poorly performing unused scenario is added together with its low-frequency mirror; at zero offset, its partner instead has both phases shifted by $\pi$. The final tail set contains 14 scenarios.

All 27 coefficients are optimized over three rounds using six initial states, $N_x=8192$, and $M=2400$. A separate guard ensemble, evaluated at $M=3200$, uses the same five offsets with phase pairs $(\pi/4,\pi/4)$ and $(5\pi/4,5\pi/4)$ to screen candidate controls for improvement of the guard objective and to limit degradation of nominal, mean, and worst-sampled fidelities.

Final evaluation of all 192 paired scenarios uses $N_x=16384$, $M=6400$, and 8 states. The quasistatic-response scan is repropagated at the same temporal resolution. For this scan, $\epsilon_\nu(t)=\delta_\nu$ and the 2- and 4-MHz components are switched off. For the 192-scenario ensemble, the independent phase coordinates are
\begin{equation}
 \varphi_j=\frac{2\pi(j+1/2)}{8},\qquad j=0,\ldots,7.
\end{equation}
Combined with three offsets, these give 192 scenarios evaluated in pairs for the nominal and robust controls. The same scenario coordinates were used in the original final acceptance evaluation at $M=3200$, which compared the aggregate objective and prescribed degradation tolerances with fallback to the nominal control. The refined $M=6400$ statistics repropagate the selected controls without reoptimization. This acceptance role must be distinguished from an untouched test ensemble.

\subsection{Noise-aware chain rule}
For a fixed scenario, the depth parameter contributes through both the commanded center and the actual depth:
\begin{equation}
\begin{aligned}
 \partial_{c_k}J_\nu
 &=\sum_\ell g_{x,\ell}^{(\nu)}\,
     \partial_{c_k}x_{0,\rm des,\ell}\\
 &\quad+\sum_\ell g_{U,\ell}^{(\nu)}(1+\epsilon_{\nu,\ell})
     \partial_{c_k}U_{\rm des,\ell}.
\end{aligned}
\end{equation}
Here $\ell$ indexes propagation time slices, $J_\nu=\ell_\nu$ is the scenario loss, and $g_{x,\ell}^{(\nu)}=\partial J_\nu/\partial x_{0,\nu,\ell}$ and $g_{U,\ell}^{(\nu)}=\partial J_\nu/\partial U_{\nu,\ell}$ are its discrete control gradients. The noise is held fixed during differentiation. There is no realization-dependent center compensation. The soft-worst gradient is a normalized exponential weighting of scenario gradients \cite{Blanchard2021}:
\begin{equation}
 \nabla\mathcal L_{\rm sw}
 =\sum_\nu\alpha_\nu\nabla\ell_\nu,\qquad
 \alpha_\nu=\frac{e^{\ell_\nu/\tau}}{\sum_\mu e^{\ell_\mu/\tau}}.
\end{equation}
The normalizing factor in Eq.~\eqref{eq:softworst} makes the smooth loss lie between the mean and maximum losses, rather than an upper bound on the maximum. For $N$ tail scenarios,
$\ell_{\max}-\tau\ln N\le\mathcal L_{\rm sw}\le\ell_{\max}$.

\FloatBarrier
\section{Control-space dimension and optimizer settings}
\label{app:basis}
Table~\ref{tab:basis} compares 21- and 27-parameter controls at all three stages. Each stored waveform is independently checked by refined spatial and temporal propagation for $L=12.6\,\mu$m and $t_f=30.5\,\mu$s, without reoptimization. The numerical settings are documented in Appendix~\ref{app:numerics}. The selected smaller-basis candidate is its depth-coupled stage; the selected larger-basis candidate is its full-joint stage. Reporting all stages keeps those selection choices distinct. The result demonstrates sensitivity to additional control freedom, rather than convergence over basis size or initialization \cite{PaganoBases2024}.
The column labels give the numbers of auxiliary-trajectory, depth, and residual-center coefficients, respectively: $8+8+5$ gives 21 parameters and $10+10+7$ gives 27. The smaller control uses auxiliary modes 1--8 and residual modes 4--8.
The $30.5\,\mu$s trajectory-only search also illustrates the distinction between a selected waveform and the best validated iterate. Its 77 accepted iterations include a cumulative minimum of about $2.33\times10^{-7}$ at iteration 56, whereas the last accepted waveform gives about $2.93\times10^{-7}$. The selected joint waveform gives about $2.41\times10^{-7}$. Thus the longer-task comparison in Fig.~\ref{fig:joint}(f) concerns the displayed selected candidates and does not establish a better global or best-achieved joint optimum. The faster-task comparison has its own search and controls.

\begin{fixedcolumntable}
\caption{Basis-dimension sensitivity. Entries are ground-state infidelities in units of $10^{-7}$. The percentage, calculated from unrounded values, is the reduction from the 21- to the 27-parameter result at the same stage.}
\label{tab:basis}
\begin{ruledtabular}
\begin{tabular}{lrrr}
Stage & $8+8+5$ & $10+10+7$ & Reduction\\
\hline
Motion & 9.40 & 2.93 & 68.8\%\\
Depth-coupled & 7.99 & 2.40 & 69.9\%\\
Full joint & 8.06 & 2.41 & 70.1\%
\end{tabular}
\end{ruledtabular}
\end{fixedcolumntable}
For the $30.5\,\mu$s ground-state search, the three stages use iteration budgets of 100, 80, and 160.

\FloatBarrier
\def\bibfont{\fontsize{9}{9.3}\selectfont}

\end{document}